\documentclass[a4paper,11pt]{article}
\pdfoutput=1

\usepackage{amsmath}
\usepackage{graphicx}
\usepackage{amsfonts}
\usepackage{amssymb}
\usepackage{floatflt}

\newtheorem{rema}{Remark}[section]

\newcommand{\bc}{\begin{center}}
\newcommand{\ec}{\end{center}}
\def\ba#1{\begin{array}{#1}\displaystyle}
\newcommand{\ea}{\end{array}}

\newcommand{\beq}{\begin{equation}}
\newcommand{\eeq}{\end{equation}}
\newcommand{\beqa}{\begin{eqnarray}}
\newcommand{\eeqa}{\end{eqnarray}}
\newcommand{\no}{\nonumber}
\newcommand{\n}{\nonumber\\}
\newcommand{\bi}{\begin{itemize}}
\newcommand{\ei}{\end{itemize}}

\def\lt#1{\left#1}
\def\rt#1{\right#1}
\def\t#1{\tilde{#1}}

\def\b#1{\bar{#1}}
\def\frc#1#2{\frac{#1}{#2}}

\newcommand{\p}{\partial}

\newcommand{\bra}{\langle}
\newcommand{\ket}{\rangle}
\newcommand{\Z}{{\mathbb{Z}}}

\newcommand{\C}{{\mathbb{C}}}
\newcommand{\hC}{{\hat{\mathbb{C}}}}
\newcommand{\uH}{{\mathbb{H}}}

\newcommand{\Or}{{\cal O}}

\newcommand{\ep}{\epsilon}

\newcommand{\id}{{\rm id}}

\newcommand{\tE}{{\tt E}}
\newcommand{\tX}{{\tt X}}

\newcommand{\tT}{{\tt T}}

\newcommand{\tI}{{\tt I}}

\newcommand{\ii}{{\rm i}}

\begin{document}

\begin{titlepage}

\begin{center}
{\Large {\bf Random loops and conformal field theory}

\vspace{1cm}

Benjamin Doyon}

Department of Mathematics, King's College London\\
Strand, London, U.K.\\
email: benjamin.doyon@kcl.ac.uk

\end{center}

\vspace{1cm}

\noindent This is a review of results obtained by the author concerning the relation between conformally invariant random loops and conformal field theory. This review also attempts to provide a physical context in which to interpret these results by making connections with aspects of the nucleation theory of phase transitions and with general properties of criticality.

\vfill

{\ }\hfill \today

\end{titlepage}

\section{Introduction}

The goal of this paper is to present a review of some of the results obtained by the author in \cite{Dcalc,DTCLE,Ddesc,Dhigher} (generalizing the earlier results \cite{FW03,DRC06}), and to put them in the context of the general theory of criticality in statistical systems.

The results reviewed concern the relation between conformal loop ensembles (CLEs) and conformal field theory (CFT)  \cite{BPZ,Gins,DFMS97}. CLEs are families of conformally invariant measures for random loops on domains of the plane \cite{Sh06,ShW07}, which emerged as generalizations of Schramm Loewner evolutions (SLEs) \cite{S00,LSW04,WSLE,KNSLE,LSLE,Cardy05,BBSLE}. SLEs, and hence CLEs, are well-known to have relations to CFT \cite{Smi1,BBSLE1,BBSLE2,BBSLE3,BBSLE4,Smi2,KNSLE,Cardy05,BBSLE,GSLE,RBGWSLE,Smi3}: they are expected to describe the universal scaling limit of a very large family of critical systems, including the universality classes of all CFT minimal models \cite{BPZ}.

CLE uses the language of probability on non-local objects, instead of that of infinite-dimensional algebras and local fields used in CFT. In this light, a CLE measure can be thought of as providing a mathematically accurate Euclidean path integral description of CFT. Although CFT models do not generically have clear Lagrangian formulations, CLE shows that nevertheless, one may obtain a measure by concentrating on non-local objects. More precisely, in a large family of two-dimensional CFT models, the scaling limits of cluster boundaries are well-defined random loops. This interpretation is particularly important, as many quantum field theory models of interest do not have an immediate Lagrangian formulation; in higher dimensional models one could expect extended objects like strings or branes to be the ones to consider. It is then essential to fully understand the physics of CLE loops and its relation to CFT.

The results obtained in \cite{DTCLE,Ddesc} relate to variables that measure shapes of loops. Exact expressions are obtained for certain correlations between such variables, and these lead to the identification of specific shape variables with the holomorphic stress-energy tensor of CFT and its descendants, and with ratios of partition functions. The mathematical tools used are studied in \cite{Dcalc,Dhigher}, and are adaptations to the present context of concepts of geometric vertex operator algebras \cite{Hu97,Hu99,Hu03}, connecting measure-theoretic, analytic and algebraic descriptions of CFT.

In the present review the CLE loop viewpoint is further connected with aspects of the theory of nucleation and with some of the fundamental tenets of criticality, like sensitivity, scale invariance and universality. The specific results of \cite{DTCLE,Ddesc} are interpreted in this context. One fundamental question that is investigated is how loop fluctuations are transferred between scales, for instance from small loops where thermal fluctuations occur, to large loops which are subject to macroscopic fluctuations. The image is similar (but perhaps inverted) to that of Richardons's cascade in three-dimensional turbulence, by which energy is transferred from large to small scales (where it is dissipated by viscosity). I obtain specific conclusions about fluctuation transfers in CLE. In particular, the central charge is interpreted as a measure of the flow of fluctuation transfer in scales, and I identify the effects on macroscopic loops of various simple types of small-loop fluctuations. This, I hope, somewhat clarifies the physical picture of CLE in the context of critical statistical mechanics.

This review is very restricted, and does not cover the large body of work, both numerical and analytical, where other connections are made between loop variables and CFT fields and states. Yet the results explained and interpreted show, I believe, that taking a different viewpoint on critical models, based on measures for non-local objects, gives a new and fertile intuition.

This paper is organized as follows. In Section 2 we provide a picture of criticality and conformal field theory from the nucleation point of view. In Section 3 we describe the shape-measuring random variable. In Section 4 we apply the latter to the construction of the stress-energy tensor. In Section 5 we extend the construction to descendants using hypotrochoids, and make a link with conformal geometry. In Section 6 we discuss possible interpretations of these results. Finally, in Section 7 we provide concluding remarks.

\section{Criticality and nucleation}

\subsection{The nucleation picture of criticality}

\begin{floatingfigure}{4cm}\bc
\includegraphics[width=4 cm]{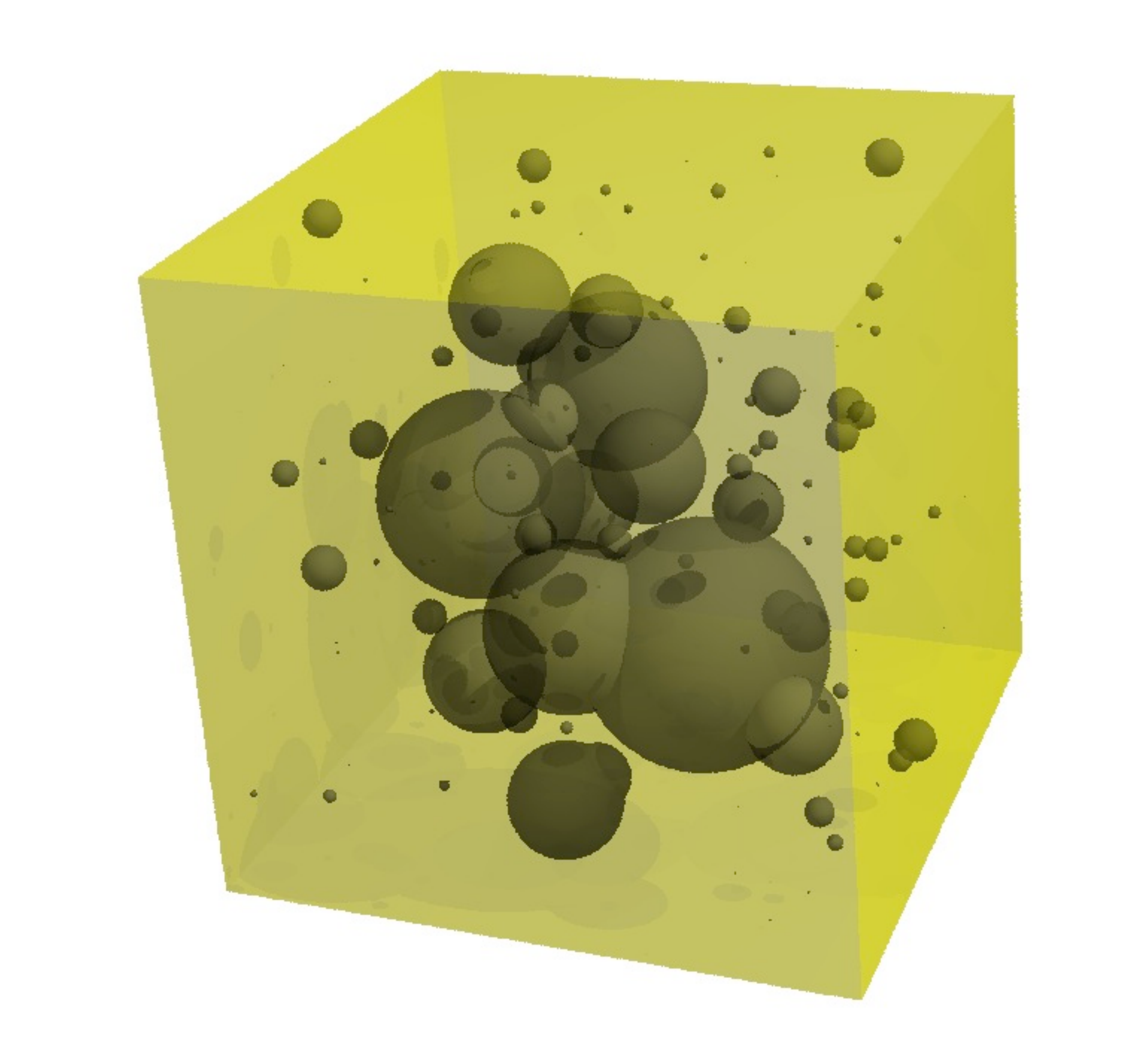}\ec
\caption{A representation of the bubbles in a critical phase.}\label{nucl}
\end{floatingfigure}
Thermodynamic systems sometimes undergo first order phase transitions characterized by finite jumps in thermodynamic properties, like the density or magnetization, upon variations of external parameters. This happens when the space of microscopic states is divided into domains, or phases, that are separated, in the topology induced by the microscopic thermal fluctuations, by regions of very low probabilities. Given some external parameters like the temperature or an external magnetic field, a particular phase dominates: ergodicity, the ability of the system to explore all microscopic states, is broken, and only the phase which minimizes the free energy is explored. As parameters change, another phase may start having smaller free energy. According to the nucleation theory (see e.g. \cite{Kal13}), a first order phase transition then occurs by nucleation of bubbles of the advantageous phase inside a bulk of the disadvantageous phase -- these bubble configurations appear to form the most likely paths between phases in the manifold of microscopic states. Bubbles are bounded by phase boundaries, which carry their own free energy cost. Hence small bubbles are suppressed, and the system is in a metastable phase. Once big enough bubbles appear, they may grow and the system may change phase. The difficulty for the system to create a big enough bubble in this mechanism of nucleation essentially represents the difficulty encountered in traversing low-probability regions when going from one phase to another.

Criticality in thermodynamic systems may be characterized by a {\em coexistence of phases}: in critical systems, the free energy is minimized when both phases coexist. Within a nucleation-theory picture, it occurs under two conditions: both phases have equal free energy, and phase boundaries carry no free energy cost\footnote{Physical phase boundaries are not codimension-1 surfaces, but rather surfaces with a thickness proportional to the correlation length. Hence at criticality, they are not defined. But one can define mathematical surfaces lying inside these physical phase boundaries \cite{Kal13}, in such a way that the above picture makes sense.}. This occurs, in parameter space, at an end-point of an arc of first-order phase transition (or possibly the equivalent with higher dimensional varieties if more parameters are present). Then, on entropy grounds, the system may advantageously produce many phase boundaries (see Figure \ref{nucl}). This restores ergodicity, and as the critical point is approached in parameter space from, say, an ordered metastable phase (with broken ergodicity), the sudden increase in entropy and decrease in free energy brought about by the apparition of many bubbles produces divergencies in thermodynamic quantities.

Exactly at the critical point, the system displays many special properties. We may identify three general properties that seem to be shared by most critical thermodynamic systems, and which find a natural understanding within the nucleation picture:
\bi
\item Sensitivity. The system is extremely sensitive to external disturbances. For instance, the application of a small external magnetic field in a magnetic critical system produces an infinite-derivative variation of the overall magnetization (the susceptibility is infinite). Within the nucleation theory, this may be understood by the fact that as soon as one phase becomes thermodynamically advantageous, it immediately nucleates to dominate the system since phase boundaries carry no free energy. Further, and associated to such divergencies of response functions, local disturbances carry effects far away, so that correlations occur also at large distances (the decay is not exponential but algebraic, the correlation length is infinite). These divergencies are characterized by critical exponents.
\item Scale invariance. The statistical properties of a critical system appear to be invariant under homotheties. One usually says that there appears to be no physical scale controlling fluctuations beyond the microscopic scales. In connection with the divergence of response functions and the existence of correlations at large distances, this led to the scaling theory of critical systems, predicting relations amongst critical exponents \cite{Fi67,Ka67}. Within the nucleation theory, one may assume that bubbles exist at all scales, uniformly in the logarithm of the scale. Large correlations may then be understood by the assumption that local disturbances, which affect the fluctuations of small bubbles, propagate easily from scale to scale up to large bubbles. The phenomenon of critical opalescence, whereby in critical systems light of all wavelengths is diffracted, may be attributed to the presence of bubbles at all scales.
\item Universality. The critical exponents observed, as well as many ratios of thermodynamic quantities and correlations, appear to be shared by large families of models. Within the nucleation theory, one may attribute this to two assumptions. First, such universal quantities are controlled by macroscopic fluctuations: local effects carry to large distances by passing through large bubbles, and large susceptibilities occur because of the formation of large bubbles. Second, the statistics of large bubbles should be universal: thermal fluctuations occur at the microscopic level, and in critical systems, they generate macroscopic fluctuations by a long chain of influence in scale, whereby an important amount of microscopic information is lost.
\ei

From the viewpoint of complex systems, large bubbles and their macroscopic fluctuations may be seen as universal emergent objects and behaviors, and it is these that lead to interesting aspects of criticality in thermodynamic systems. One is then led to investigate how to describe these universal objects, by-passing the complicated and irrelevant microscopic depiction. In particular, the above three properties should be naturally embedded in such a description.

A concept unifying these three properties is that of fluctuation transfer from small to large scales. As we mentioned, within the nucleation picture we may see such transfers as instrumental in explaining large susceptibilities and large-distance correlations. A constant transfer over large ``scale distances'' is also synonymous to scale invariance, and the fact that only some information is not ``washed out'' over such large scale distances is universality. It is worth noting the connection with turbulence and Richardson's energy cascade amongst scales, with the viscosity playing the role of a microscopic cutoff to this process. In any case, in the context of statistical mechanics, one may then postulate the existence of a number $c$ which characterizes the quantity of universal fluctuations transferred between scales, or some bigger structure characterizing more intricate aspects of fluctuation transfers. This and the above nucleation picture of criticality can be made somewhat more precise in {\em two-dimensional statistical systems}.

\subsection{Two dimensions: spins and random loops}

A paradigmatic model of statistical mechanics is the Ising model, solved by Onsager in 1944 \cite{Onsa}. It is a model where configurations are spins, taking values 1 or -1, lying on sites of a lattice $L$. Spins interact with each other according to the energy functional $H = J\sum_{(i,j)} \sigma_i \sigma_j - h\sum_i \sigma_i$, so that the measure is $\mu = e^{-H/T}$ where $T$ is the temperature. Here $(i,j)$ are pairs of neighbours on the lattice, and $\{\sigma_i:i\in L\}$ is a spin configuration. The quantities $J$ and $h$ are the interaction energy and external magnetic field, respectively. Although the language used here is that of a magnetic model, the Ising model has been applied to many other situations.

Let us consider $J=-1$, where the model is ferromagnetic. When the temperature is smaller than a critical temperature $T_c$ (the exact value of which depends on the lattice), the model displays two phases: one with a positive magnetization (positive average spin at any site), another with a negative magnetization. For positive (resp. negative) magnetic field $h$, the former (resp. latter) phase is advantageous. As the magnetic field values are continuously changed from positive to negative while the system is in a positive-magnetization phase, the system undergoes a first order phase transition: bubbles, here loops as we are in two dimensions, of the negative-magnetization phase are created, and eventually one grows to dominate the whole system. At the temperature $T_c$, the free energy cost associated to a boundary between the phases is zero, and the system is critical at $T=T_c,\,h=0$.

\begin{floatingfigure}{4cm}\bc
\includegraphics[width=4 cm]{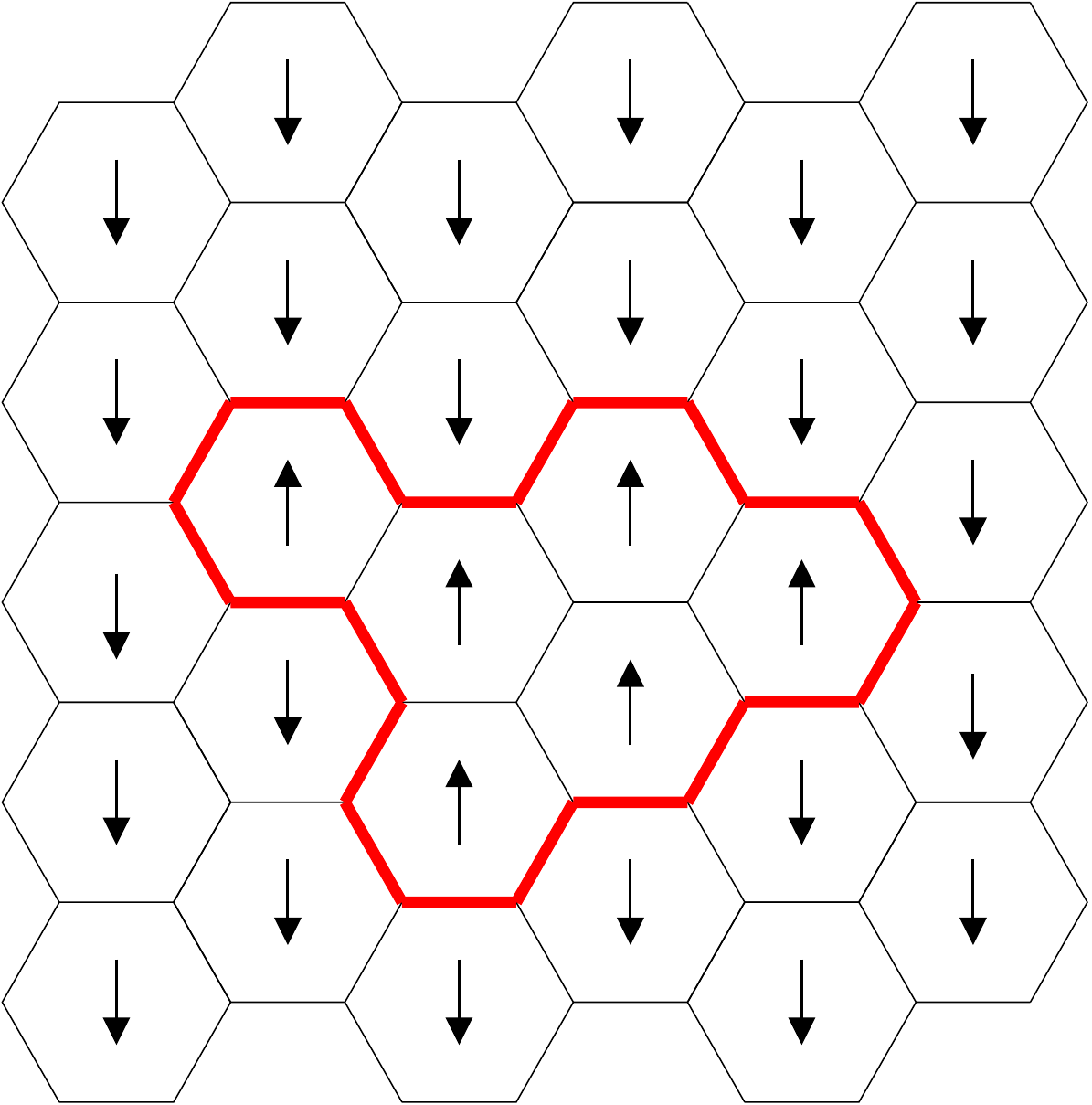}\ec
\caption{A cluster boundary.}\label{latloops}
\end{floatingfigure}

In general, the precise definition of a phase boundary is not unique. In the present case, however, there is a natural definition. For definiteness, let us assume that the spins lie on the faces of an infinite honeycomb lattice. Consider the set of all  edges separating positive- and negative-spin neighbors. With appropriate asymptotic conditions, this becomes a set of closed disjoint simple loops (see Figure \ref{latloops}). In fact, such a set of loops, along with the specification of the phase, is an equivalent description of the configuration\footnote{Specifying the phase, positive or negative magnetization, is needed because the set of loops is unchanged under a change of sign of all spins.}. In the Ising model, the loops are where the positive energy contributions to the interaction term lie, as it is where the term $J\sigma_i \sigma_j$ is positive; under a constant shift of the energy functional, the loops can be seen as the locus of the interaction energy.

One may then define a phase boundary as a {\em macroscopic loop} in this description. That a loop be associated with a phase boundary is intuitively clear, as it separates between positive and negative spins. However, in any phase, due to thermal fluctuations, loops are always present -- they cannot all be identified with phase boundaries. But such ``fluctuation loops'' are almost surely microscopic (more precisely, of the size of the correlation length or less). Loops that can be associated with the presence of a phase boundary are the macroscopic ones; for instance, when a bubble grows according to the nucleation theory, the loop does become macroscopic.

This definition agrees with the description of criticality that we made in the previous section. Indeed, at critical points it is observed that there are almost surely infinitely many macroscopic loops. Further, intuitively, macroscopic loops not only carry a positive energy cost $U$ (the associated interaction energy), but also an entropy $S$, because of the many microscopic configurations leading to a given macroscopic shape. One may then in principle define a free energy cost $U-TS$ associated to the loop, and it is this free energy cost that tends to zero at criticality (but we do not attempt to develop this idea here). Then, at criticality, macroscopic loops should be very ``wiggly'' in order to accommodate a large entropy and make their free energy cost zero; this is indeed what is observed.

\begin{floatingfigure}{4cm}\bc
\includegraphics[width=4 cm]{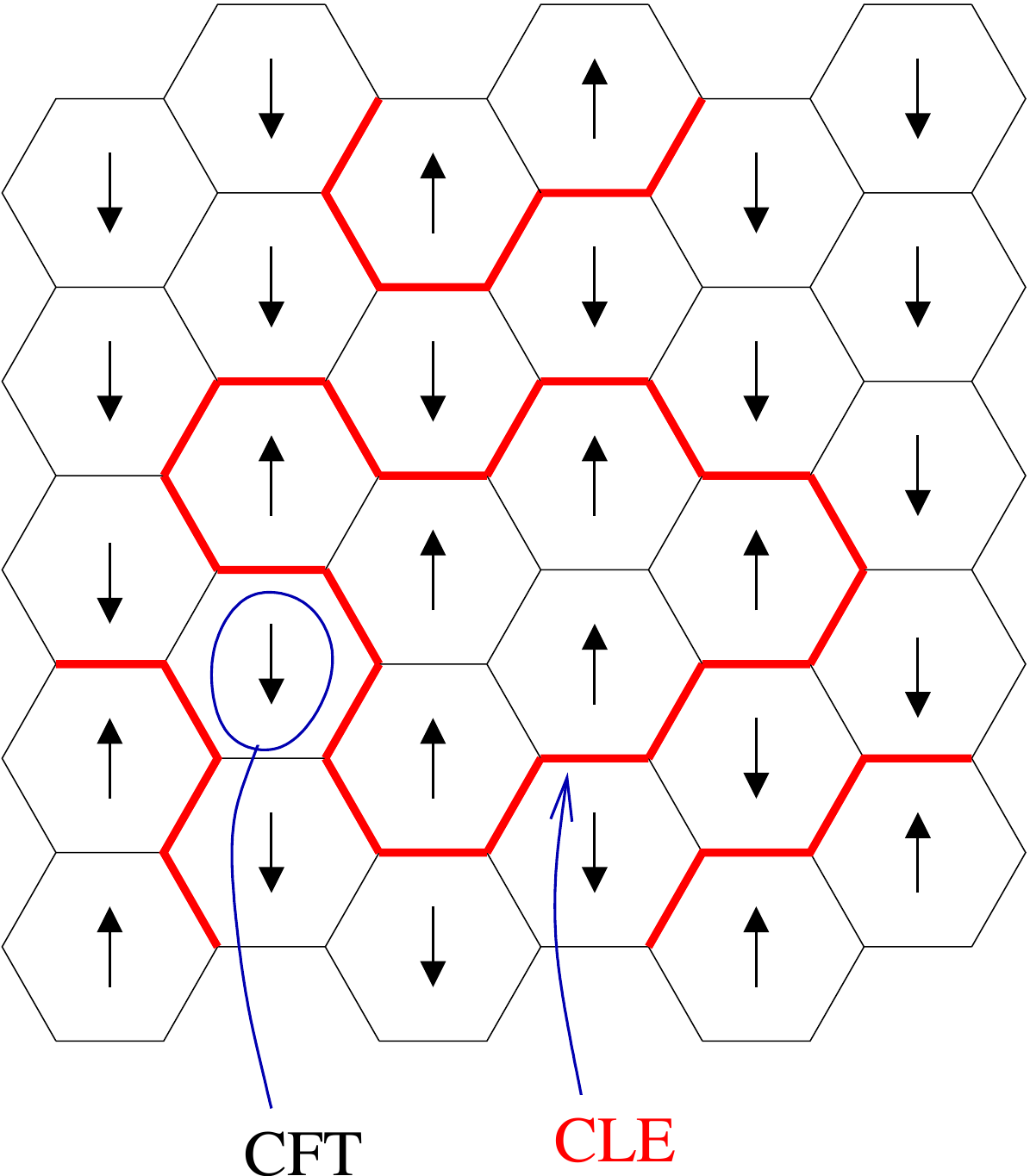}\ec
\caption{What CFT and CLE describe.}\label{CFTCLE}
\end{floatingfigure}

The exact, mathematically rigorous description of macroscopic loops in the critical Ising model is that given by conformal loop ensembles (CLE) \cite{Sh06,ShW07}  (see Figure \ref{CFTCLE}). This is a family of measures for random loop configurations, which is expected to correspond to a one-parameter family of universality classes of critical two-dimensional statistical models, including the Ising model. In particular, a family of microscopic models whose scaling limit is expected to be CLE is given by the so-called $O(n)$ models, parametrized by a real parameter $n$. These are models for random loops on the edges of (say) the honeycomb lattice. One can express the energy functional by $H = \ell + m \,{\cal N} $ where $\ell$ is the total number of occupied edges, ${\cal N}$ is the total number of loops, and $m$ is a parameter (the case $m=0$ is the Ising model). With the usual thermal measure $\mu = e^{-H/T}$ at temperature $T$, and defining $x:=e^{-1/T}$, the system is critical at $x = x_c$ satisfying $x_c \,\sqrt{2+\sqrt{2-x_c^m}} = 1$ \cite{N82}. The more commonly used parameter $n$ of the $O(n)$ model is related to $m$ by $n = e^{-m/T}$.

We may see the CLE description of critical points in terms of their phase boundaries as a mathematically precise ``nucleation picture''. It is interesting to enquire if the expected properties of criticality (sensitivity, scale invariance and universality) can be understood in more depth within it. For instance, we would like to gain intuition concerning the mechanisms for transfer of fluctuations from small to large scales: is there a quantity or mathematical setup which characterizes this transfer? We will review some exact results obtained by the author in the context of CLE, and provide an interpretation in this direction.

\subsection{Conformal field theory and conformal loop ensembles}

Conformal field theory (CFT) provides exact conjectures for the scaling limit of correlation functions of local observables in critical statistical systems (see Figure \ref{CFTCLE}). It can be seen as an algebraic framework, based on the Virasoro algebra (and on vertex operator algebras), which allows one to give predictions for the way correlation functions decay as the positions of local observables are scaled up; or, equivalently, as the mesh size of the lattice is decreased, while observables' positions are fixed with respect to the background. For instance, denoting by $\mathbb{E}_\lambda(\cdots)_A$ the expectation value of the Ising model on a lattice of mesh size $\lambda$ inscribed in a domain $A$, we have
\beq\label{scalinglim}
	\lim_{\lambda\to0} \lambda^{-nd} \mathbb{E}_\lambda (\sigma_{[x_1]}\cdots \sigma_{[x_n]})_{A} = \bra\sigma(x_1)\cdots\sigma(x_n)\ket_A
\eeq
where $x_i$ are positions of the spin variables, $[x_i]$ are the lattice sites nearest to $x_i$, and $\bra\cdots\ket_A$ is the CFT correlation function. In this case it is known that $d=1/8$ and that the local field $\sigma(x)$ is a Virasoro primary field \cite{BPZ}. The CFT correlation function up to an overall normalization of the fields, and the number $d$, are universal quantities.

CFT provides a classification of local fields based on the Virasoro algebra. However, this is {\em a priori} difficult to link with local random variables in lattice models in the sense of \eqref{scalinglim}, although much progress has been made in this direction recently with rigorous results in the Ising case \cite{Smi3,HS10,CHI12,HKZ12}. This difficulty is particularly true for the stress-energy tensor $T(w)$: it is expected to possess a wealth of important properties based on fundamental QFT principles, yet it is not easily identifiable as a local variable on the lattice.

\begin{floatingfigure}{4cm}\bc
\includegraphics[width=4 cm]{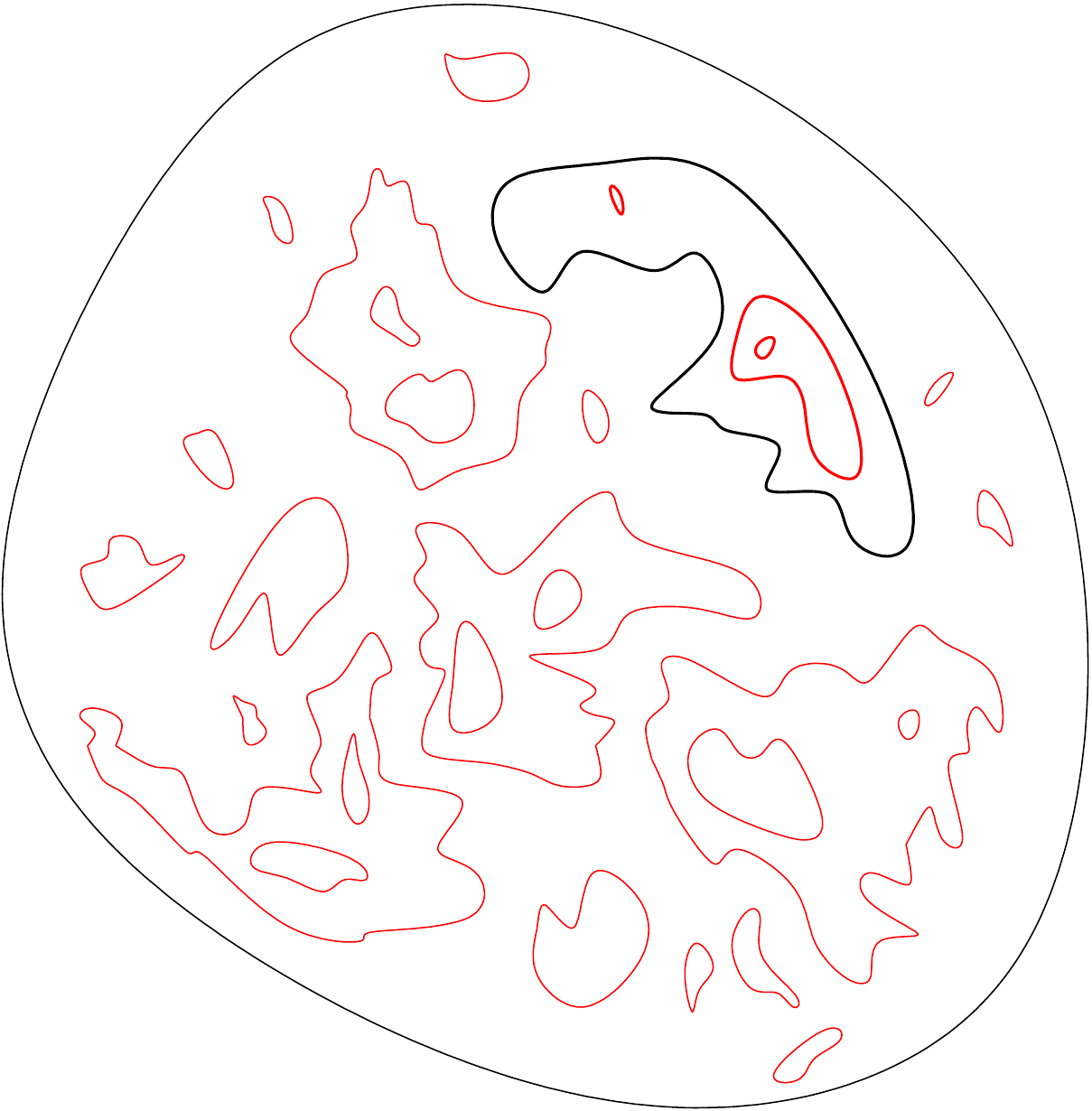}
\includegraphics[width=4 cm]{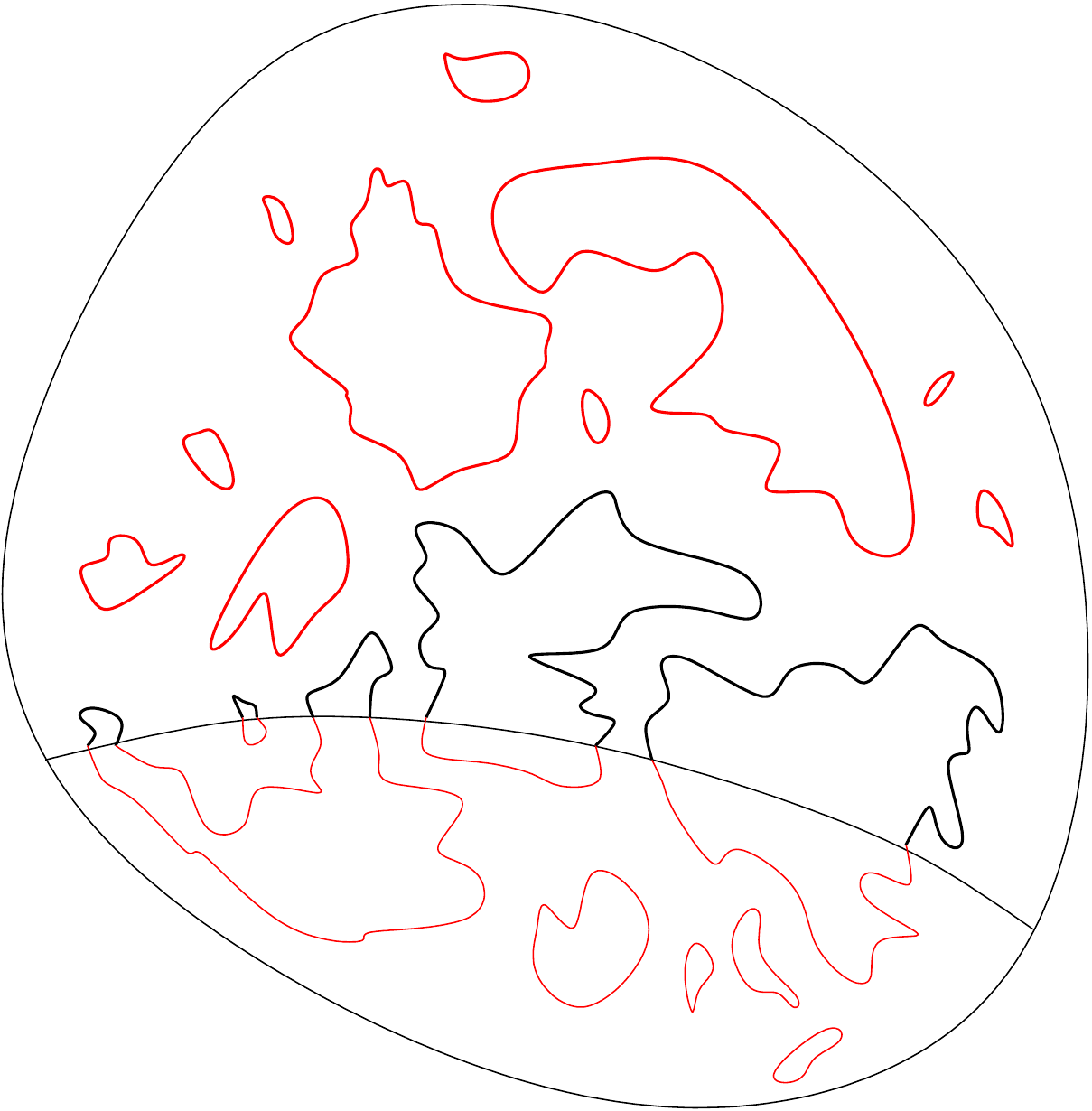}\ec
\caption{Nesting and conformal restriction.}\label{inout}
\end{floatingfigure}

In two-dimensional CFT,  thanks to Noether's theorem, translation, rotation and scale invariance imply the existence of two local fields $T(w)$ and $\b T(\b w)$ which are, respectively, holomorphic and anti-holomorphic functions of their position. The fact that these generate translations, rotations and homotheties translates into conditions $\oint_z dw\, (w-z)^{m+1} T(w) \Or(z) = \delta_m \Or(z)$ relating holomorphic contour integrals encircling fields $\Or(z)$ with the infinitesimal field transformation $\delta_m \Or(z)$, for $m=-1,0$. Clearly, the result must be a local field at $z$ for any value of $m$. Further, by dimensional analysis, it must be a linear combination of fields of equal or lower scaling dimensions, for any $m\geq 0$ (for $m=-1$, only derivatives can be involved by Noether's theorem for translation invariance); and with the condition that the dimensions of fields are bounded from below, the result must be zero for all $m$ large enough. Hence, holomorphicity of $T(w)$ essentially implies {\em local conformal invariance}: there are transformation properties $\delta_m \Or(z)$ for all $m\geq 0$ involving equal- and lower-dimension fields, and there are fields for which $\delta_m\Or(z)=0$ for all $m\geq 1$ (primary fields). This includes the case $m=1$, generating special conformal transformations. Taking $\Or(z) = T(z)$, one obtains the Virasoro algebra for the generators $\oint_z dw\, (w-z)^{m+1} T(w)$. The value of the central charge $c$ of the Virasoro algebra is a characteristic of the model under consideration. These infinitesimal transformations exponentiate to conformal transformations, which then give rise to conformal invariance equations for correlation functions. Although there are infinitely many $m$, given a domain $A$, there is only a finite-dimensional group of conformal transformations (which we always understand as bijective) that preserve $A$. Hence, the conformal symmetry, on a given domain, is {\em finite-dimensional}. Yet, local conformal invariance, often referred to as infinite-dimensional conformal symmetry, is extremely powerful.

there are only finitely many conformal transformations that preserve $A$. Hence, the conformal symmetry, on a given domain, is {\em finite}. Yet, local conformal invariance, often referred to as infinite conformal symmetry, is extremely powerful.

The above considerations, on the other hand, are not easily interpreted in the context of CLE. Indeed, CLE measures $\mu_A$ describe the behaviors of macroscopic loops instead of local fields: a CLE configuration on a domain $A$ of the Riemann sphere is a set of non-crossing and non-self-crossing loops in $A$. One parametrizes the CLE measures by $\kappa\in[8/3,8]$, which indicates how dense and how ``wiggly'' loops are. In particular, it is expected that the central charge $c$ of CFT be related to the parameter $\kappa$ of CLE by \cite{BBSLE1,Sh06,ShW07}
\beq\label{ckappa}
	c = \frc{(3\kappa-8)(6-\kappa)}{2\kappa}.
\eeq
For $\kappa>4$, the density is such that loops have double points and have common points with each other and with the boundary of the domain; for $\kappa\leq 4$, loops are simple and disjoint from each other and from the boundary of the domain. At $\kappa=8/3$ ($c=0$), there are no loops at all (the theory is trivial -- although it is possible to make it nontrivial by adding other types of curves). Loops have fractal dimension $1+\kappa/8$ \cite{Bef08}, and in every configuration there are infinitely many loops, although there are always finitely many loops with a diameter greater than any given positive number. In the following, we consider only the dilute phase, $\kappa\leq 4$. In this case, the family of measures, when restricting to simply connected domains, is completely determined by three defining conditions: {\em conformal invariance}, $\mu_{g(A)}\circ g = \mu_A$; {\em nesting}, the measure on all loops lying inside a loop is that of the CLE on the domain bounded by that loop; and {\em (probabilistic) conformal restriction}, something similar, but from the viewpoint of what lies outside a random domain formed by the union of a fixed domain and the interiors of all loops that intersect it (see Figure \ref{inout}). CLE in the dilute phase was mathematically proven to exist and explicitly constructed in \cite{ShW07}.

The main goal of the works \cite{Dcalc,DTCLE,Ddesc,Dhigher} whose results we review here, was to provide a bridge between CFT correlation functions and CLE expectation values. This, in particular, gave the solution as to what the stress-energy tensor is in terms of CLE loops. Note that in CLE there is no explicit notion of ``local conformal invariance'' -- there is no immediate notion of local fields, and no property that allows one to consider transformations that are conformal only locally, as one has to deal with the full domain of definition. We believe the construction of the stress-energy tensor in CLE makes clearer what local conformal invariance actually stands for.

\section{Measuring the shape of small loops}

\begin{floatingfigure}{2cm}\bc
\includegraphics[width=2cm]{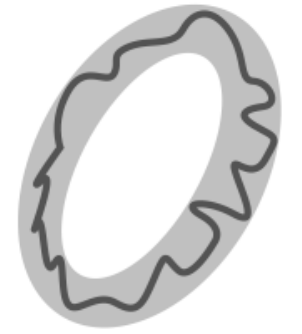} \ec
\caption{$\tI(N)$.}\label{IN}
\end{floatingfigure}

We discussed above how there should be a quantity measuring the transfer of fluctuations between scales; one may guess that this could be the central charge $c$  of CFT. Can we make this statement more precise? That is, can we relate $c$ to fluctuations of small loops and their effects on large scales or their interaction at large distances? In order to study loop fluctuations, let us consider the natural question of measuring the shape of loops: what is the probability for a loop to be of a given shape?  

\subsection{A variable for the shape of loops}


Consider the indicator variable $\tI(N)$ that indicates if (i.e. is one if and zero otherwise) at least one CLE loop winds in an annular domain $N$ (see Figure \ref{IN}). In order to measure the shape of loops, we would like to take the limit where the annular domain $N$ becomes very thin and tends to a fixed shape (loop) $\alpha$ (see Figure \ref{Ntoalpha}). Of course, in this limit the expectation of the indicator variable $\tI(N)$ tends to zero: there is zero probability that at least one loop takes a given fixed shape. This is clear for the simple reason that loops can take any shape in a continuous way -- any point on this continuous space of shape naturally has probability zero. But also, loops are very ``wiggly", with a fractal dimension related to the central charge. This makes the expectation of $\tI(N)$ decay even faster than otherwise as $N\to\alpha$.

\begin{floatingfigure}{4cm}\bc
\includegraphics[width=4 cm]{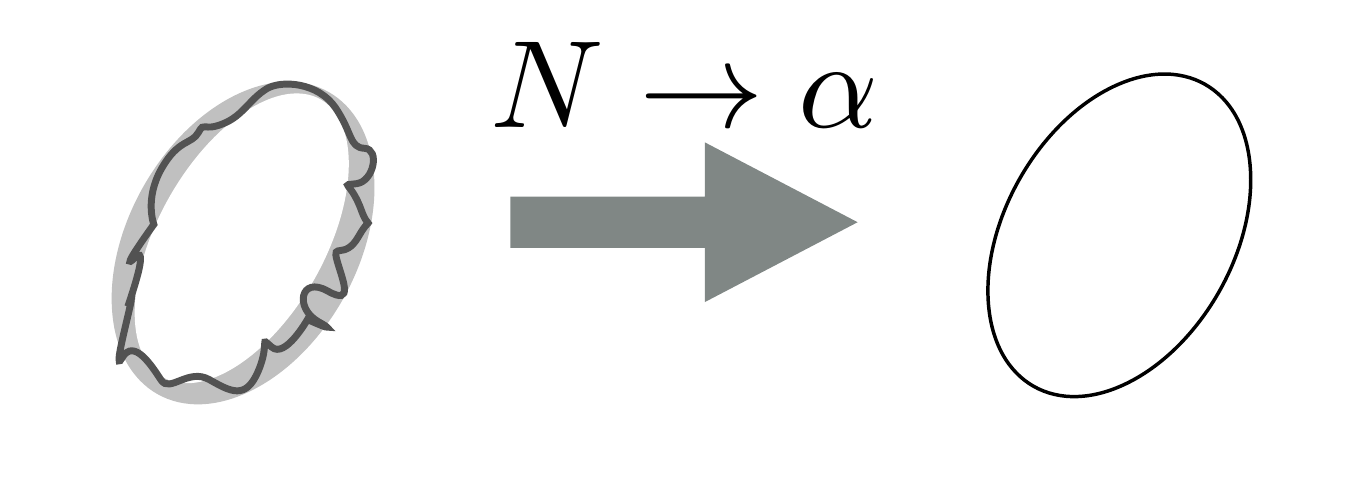} \ec
\caption{$N\to\alpha$.}\label{Ntoalpha}
\end{floatingfigure}

Nevertheless, we may attempt to normalize the variable, in order that the limit be something nonzero:
\beq\label{tE}
	\tE(\alpha):=\lim_{N\to\alpha} \frc{\tI(N)}{\mathbb{E}\big[\tI(N)\big]_{\hC}}.
\eeq
Here we have normalized by dividing by the expectation of $\tI(N)$ on a fixed, arbitrary domain which we have chosen to be the Riemann sphere $\hC$. Note that we don't know explicitly what normalization factor to take so that the limit be generically finite and nonzero, but we do expect that the expectation value $\mathbb{E}\big[\tI(N)\big]_{\hC}$ will fulfill this requirement. Equation \eqref{tE} is a type of {\em renormalization process}: we normalize then take a limit. The result is a variable that measures the weight of loops that are infinitesimally ``near'' the shape $\alpha$.

\begin{rema}
In order to make a connection with objects well known to physicists in gauge theory, we may think of $\tE(\alpha)$ as being related to {\em Wilson loops}. If we imagine that the random loops of our model come from the level lines of a fluctuating height field $\varphi$, then the current associated with the internal symmetry under height shifts is the divergence of the height fields $\p_\mu\varphi$. Let us instead think of the angle $\theta$ of the local height lines with respect to the horizontal axis. The divergence of this field, $\p_\mu\theta$, can be seen as a dual current. Since $\tE(\alpha)$ requires a loop to be of the shape $\alpha$, then the holonomy of $\p_\mu\theta$ along $\alpha$ is $2\pi$. Hence, our ``Wilson loop'' variable associates to a $2\pi$ holonomy of the dual current the value 1, and, essentially, to other holonomies the value 0.
\end{rema}

The renormalization process in \eqref{tE} is crucial in order to obtain a finite variable associated to a condition for loops to be of a fixed shape $\alpha$. Yet the existence of a finite limit in \eqref{tE} is rather nontrivial, and still not proven in the context of CLE. However, we may expect that the right hand side of \eqref{tE} will exist and be generically finite and nonzero whenever it is evaluated inside expectation values, in conjunction with any other variables that ``lie away'' from $\alpha$, intuitively variables that do not affect directly what's happening on the fixed loop $\alpha$.

This, in fact, forces us to consider the notion of where a random variable ``lies'': we need to know what region of space is affected by the insertion of this random variable in an expectation value. A natural concept is that of the support of the random variable \cite{DTCLE,Ddesc}: this is a closed set on the Riemann sphere, which is such that if one knows, in a given random loops configuration, exactly all loops that intersect this closed set, but none of the other loops, then one has a sufficient amount of information to determine the value of the random variable\footnote{This is not unique and the proper concept is for the support to be a set of such closed sets of the Riemann sphere.}. This gives some locality structure to the random variables: we know where they lie. Then, we expect the limit in \eqref{tE} to exist inside expectation values with insertion of any other random variable that lies on closed sets not intersecting $\alpha$. This is ``weak local'' convergence (weak because it is inside expectation values, local because of the additional locality requirement). This also tells us that the renormalized variable $\tE(\alpha)$ itself can be seen as lying (being supported) on the loop $\alpha$.

Finally, it is important to mention the effect of the renormalization on the transformation properties of the variable $\tE(\alpha)$. Given any conformal map $g$ acting on a domain that includes the annular domain $N$, it is natural to define its action on the random variable $\tI(N)$ by
\beq\label{gI}
	g\cdot \tI(N) = \tI(g(N)).
\eeq
This is a good action in the usual sense (that is, $g_1\cdot (g_2 \cdot \tI(N)) = (g_1\circ g_2)\cdot\tI(N)$), and it also guarantees that expectation values are conformally invariant thanks to conformal invariance of the CLE measure:
\beq
	\mathbb{E}\big[\tI(N)\big]_A = 
	\mathbb{E}\big[g\cdot \tI(N)\big]_{g(A)},\quad g\mbox{ conformal on }A,\quad N\subset A. 
\eeq

However, the transformation property of the variable $\tE(\alpha)$ is not as simple. Indeed, if $g$ is conformal on $N$ but does not preserve the Riemann sphere, then the denominator in \eqref{tE} is not invariant under $N\mapsto g(N)$. Rather, by a small calculation one finds
\beq\label{transE}
	g\cdot \tE(\alpha) = \frc{\tE(g(\alpha))}{F(g,\alpha)},\quad
	F(g,\alpha) := \lim_{N\to\alpha} \frc{\mathbb{E}\big[\tI(N)\big]_{\hC}}{\mathbb{E}\big[\tI(g(N))\big]_{\hC}}.
\eeq
That is, the transformation property of the renormalized variable gets an extra multiplicative factor that depends on $\alpha$ and $g$. This can be seen as the ``breaking'' of local conformal invariance due to a renormalization process coming from a symmetry-breaking regularization. It is just local conformal invariance that is broken, because if $g:\hC\to\hC$, then certainly the denominator in the definition of $F(g,\alpha)$ is equal, by conformal invariance of CLE, to the numerator:
\beq\label{Fmob}
	F(G,\alpha) = 1,\quad G \mbox{ a M\"obius map}.
\eeq
Hence indeed, global conformal invariance is preserve. These are important concepts behind the appearance of the Virasoro algebra with nonzero central charge in CLE.

\subsection{Shape correlation of small rotating ellipses} \label{ssectshapecorr}

\begin{floatingfigure}{3cm}\bc
\includegraphics[width=1.6 cm]{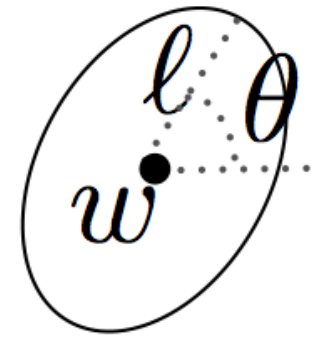} \ec
\caption{The ellipse $\alpha(w,\theta,\ell,e)$.}\label{ellipse}
\end{floatingfigure}

We now specialize the shape to a very simple one: an ellipse. Take $\alpha = \alpha(w,\theta,\ell,e)$ to be an ellipse centered at the point $w\in \C$, at an angle $\theta$ from the positive real direction, and with major semi-axis of length $\ell$ and eccentricity $e$ (see Figure \ref{ellipse}). There is no exact analytic expression yet for the expectation values and correlations of the variables $\tE(\alpha)$ with this or any other shape. However, if we make the ellipse very small and make it ``rotate'', then we do have exact results.

Let us first make this ellipse rotate with a spin 2 -- there is no dynamics here of course, but what we mean is that we are taking the second Fourier mode of the variable $\tE(\alpha(w,\theta,\ell,e))$ as a function of the angle $\theta$:
\beq\label{intE}
	\int_0^{2\pi} d\theta\,e^{-2\ii\theta} \,\tE(\alpha(w,\theta,\ell,e)).
\eeq
Then, let us take the limit where $\ell$ becomes zero. Again, we do that weakly locally: inside expectation values with possible other variables lying away from $w$. Naturally, in this limit, any domain boundary or random variable that lies away from the point $w$ (in some expectation value) will not affect the loops that are being measured by \eqref{intE}. Indeed, the variable $\tE(\alpha(w,\theta,\ell,e))$ will only be affected by very small loops near to $w$, and these small loops are shielded by the infinitely many loops that there almost surely are around any point in CLE. So, in this limit, the variable $\tE(\alpha(w,\theta,\ell,e))$ will feel as if nothing else were present on the whole plane and hence will be $\theta$ independent. Therefore, in the weak local sense, the second Fourier mode should be zero. This is indeed the case, but it turns out that we know exactly how it tends to zero \cite{DTCLE}: it does it like $\ell^2$. Whence we can normalize it so that the resulting limit $\ell\to0$ is generically nonzero. That is, we define
\beq\label{T21}
	\tT(w) = \lim_{\ell\to0} \frc1{2\pi \ep^2}\int_0^{2\pi} d\theta\,e^{-2\ii\theta} \,\tE(\alpha(w,\theta,\ell,e)),\quad \ep = \frc{e\ell}2
\eeq
as a weak-local limit, and $\tT(w)$ is then a new renormalized variable lying (supported) on the point $w$. We note that ${\tt T}(w)$ is not invariant under rotations and scaling: we see from its definition and using \eqref{transE} and \eqref{Fmob} that it has spin 2 and scaling dimension 2.

Then, one result of \cite{DTCLE,Ddesc} is that
\beq\label{2pt}
	\mathbb{E}\big[\tT(w_1)\tT(w_2)\big]_{\hC} =
	\frc{c/2}{(w_1-w_2)^4}
\eeq
where $c$ is the central charge of the CFT model expectedly associated to the CLE measure, given by \eqref{ckappa}. We notice of course that \eqref{2pt} is nothing else but the two-point function of the stress-energy tensor on the plane in CFT, $\bra T(w_1)T(w_2)\ket_\hC$. That is, it looks as though we can identify $\tT(w)$ in CLE with the stress-energy tensor $T(w)$ in CFT.

Before making this identification completely accurate, let us discuss the meaning of our result \eqref{2pt} in terms of loop fluctuations. We first note that in regularizing the limit $\ell\to0$ by dividing by $\ell^2$, so that the limit exist and be nonzero, we are in a sense focussing on the small effects that large scale objects, like domain boundaries and random variables lying away from $w$, have on the probabilities of finding a small loop with the elliptical shape at various angles. Indeed, we say that the limit exists weakly locally as an object supported on $w$, hence that it exists when inside expectation values on any domains containing $w$, and with insertions of other variables lying away from $w$. Expectation values of $\tT(w)$ then measure certain kinds of correlations between small-scale shapes and large-scale structures.

\begin{floatingfigure}{3cm}\bc
\includegraphics[width=3 cm]{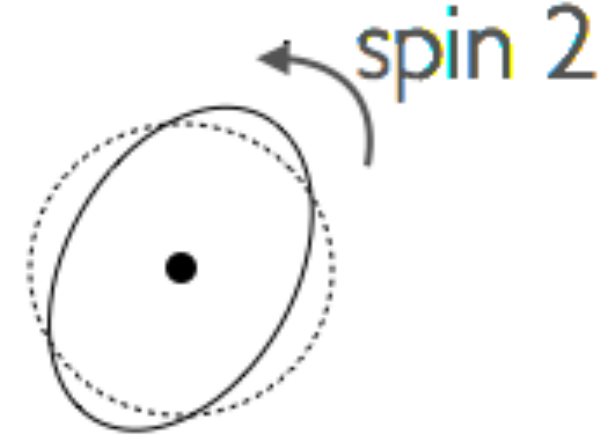} \ec
\caption{Two-crest wave propagating at spin 2.}\label{spin2}
\end{floatingfigure}

We will come back to this aspect below, but we note for now that the explicit result \eqref{2pt} does not include any domain boundary or other random variables than the two variables $\tT(w_1)$ and $\tT(w_2)$. What it does is to measure the small correlations shape fluctuations of small loops may have amongst each other at large distances -- Equation \eqref{2pt} is possibly the simplest {\em shape correlation}. We can in fact be a bit more precise about our interpretation. The spin-2 rotating ellipse can be interpreted as a two-crest wave ``propagating'' along a circle at a spin-2 speed (see Figure \ref{spin2}). Hence, Equation \eqref{2pt} tells us about the correlation these particular types of small loop fluctuations -- 2-crest, spin-2 waves -- have at large distances. Since small loops do not directly affect each other, as they are shielded by very large numbers of loops surrounding them, this correlation must come from a transfer of fluctuations scale by scale, between the small scales around $w_1$ and $w_2$ and the large-scale loops of diameter of order $|w_1-w_2|$ that surround both points. Hence, the CFT central charge $c$ may be seen as a measure of the intensity of this transfer of fluctuations.

\section{Ellipses and the CFT stress-energy tensor}

The result \eqref{2pt} is just one amongst a large number of results that unambiguously identify the variable $\tT(w)$ in CLE with the CFT (holomorphic) stress-energy tensor.

\subsection{Basic properties of $\tT(w)$}

Correlation functions on $\hC$ with more than two stress-energy tensor insertions have explicit expressions in CFT that can be obtained by recursively using the conformal Ward identities. This was found \cite{DTCLE,Ddesc} to hold as well for $\tT(w)$ in CLE:
\beqa\label{Tmulti}
	\lefteqn{\mathbb{E}\big[\prod_{k=1}^n \tT(w_k)\big]_{\hC}} \\
	&=& \sum_{j=2}^n \frc{c/2}{(w_1-w_j)^4}
	\mathbb{E}\big[\prod_{k=2\atop (k\neq j)}^n \tT(w_k)\big]_{\hC}
	+ \sum_{j=2}^n
	\lt(\frc{2}{(w_1-w_j)^2} + \frc{1}{w_1-w_j} \frc{\p}{\p w_j}\rt)
	\mathbb{E}\big[\prod_{k=2}^n \tT(w_k)\big]_{\hC}. \no
\eeqa
Note of course that $\mathbb{E}\big[\tT(w)\big]_\hC=0$ by rotation covariance. This gives exactly all higher correlation functions of $\tT(w)$ in $\hC$.

The above in fact indicates that $\tT(w)$ seems to have a particular analytic structure, as does the CFT stress-energy tensor. Indeed, it was shown that $\tT(w)$ is, weakly locally, an analytic function of $w$, and that, again weakly locally, Wilson's operator product expansion (OPE) of the CFT stress-energy tensor holds for $\tT(w)$ as well:
\beq\label{TOPE}
	\tT(w_1)\tT(w_2) = \frc{c/2}{(w_1-w_2)^4} + \frc{2}{(w_1-w_2)^2}
	\tT(w_2) + \frc{1}{w_1-w_2} \frc{\p}{\p w_2} \tT(w_2)
	 + \ldots
\eeq
where the ellipsis ``$\ldots$" corresponds to a series in nonegative integer powers in $w_1-w_2$. The combination of this singularity structure along with the elsewhere-analytic requirement on $\hC$ gives rise to \eqref{Tmulti} thanks to Liouville's theorem. But the OPE \eqref{TOPE} holds much more generally than in multi-point correlation functions of the stress-energy tensor on the Riemann sphere; it holds weakly locally: on any domain and with any other insertions away form $w_2$. Relation \eqref{TOPE} is just a particular case of the conformal Ward identities. The full series does not have a nonzero weak-local radius of convergence (that is, a nonzero radius of convergence that would be correct completely generally in the weak-local sense). Rather, inside any expectation with other variables supported away from $w_2$, the radius of convergence is given by the Euclidean distance from $w_2$ to the nearest insertion.

The relation \eqref{TOPE} was in fact not proven directly, but rather deduced from the transformation properties of the variable $\tT(w)$ \footnote{This is the opposite direction from what is usually taught in CFT: one usually derives \eqref{TOPE} from a local infinitesimal transformation analysis, with the stress-energy tensor being the generator, and extrapolates to finite transformations.}. That is, it was shown \cite{DTCLE}, from \eqref{transE} and an analysis of the factor $F(g,\alpha)$, that for any $g$ conformal on a neighborhood of $w$,
\beq\label{transT}
	g\cdot \tT(w) = (\p g(w))^2 \,\tT(w) + \frc{c}{12} \{g,w\}
\eeq
where $\{g,w\}$ is the Schwartzian derivative of $g$ at the point $w$. By our initial definition \eqref{gI}, the $g$-action is a symmetry transformation. That is, let $\tX$ be any product of variables $\tT(w)$'s at different points $w$'s, and of variables $\tI(N)$'s on disjoint annular regions $N$'s not intersecting the points $w$'s (or any other CLE variables supported away from the $w$'s). Denoting by $g\cdot \tX$ the result of acting with $g$ on each factor as per \eqref{transT} or \eqref{gI} (or in the appropriate way for other CLE variables), then
\beq
	\mathbb{E}\big[g\cdot \tX\big]_{g(A)}
	= \mathbb{E}\big[\tX\big]_{A}
\eeq
for every $g$ conformal on $A$ (where $A$ is a domain that must include the support of $\tX$). This is of course just the usual conformal invariance of CFT, with the appropriate conformal transformation property \eqref{transT} of the stress-energy tensor.

\subsection{Extended conformal Ward identities}

As is well known, one can generalize the above, in the CFT context, to a relation between the singular part of the OPEs of the stress-energy tensor  with any other field, and the transformation properties of this other field. This generalization is the expression of the general conformal Ward identities in CFT. Evaluating correlation functions in CFT as analytic functions of the stress-energy tensor position $w$ necessitates more than the singular part of the OPEs (the positions of singularities in $w$): one also needs the boundary conditions on the domain boundaries, if there are any (on $\hC$ there aren't and one doesn't need that, because one can use Liouville's theorem as above). It was understood in \cite{Dcalc} that the combination of the singular part of the OPEs (i.e. the conformal Ward identities) with the boundary conditions can be re-expressed in a compact way as a single relation referred to as the {\em extended conformal Ward identities} (not to be confused with Ward identities coming from extended symmetries like $W$-algebras). This is simply achieved by identifying the insertion of the stress-energy tensor at the point $w$ with the result of a small (infinitesimal) conformal transformation that is {\em singular} (i.e. non-conformal) around the point $w$ (in an infinitesimally small region).

The conformal transformation to use is ($\eta\in\C$)
\beq\label{gweta}
	g_{w,\eta}(z) = z + \frc{\eta}{w-z},
\eeq
which is a Joukowsky transformation, mapping the outside of a small disk centered at $w$ to the outside of a small ellipse centered at $w$. Another ingredient necessary to express the extended conformal Ward identities is  the {\em relative partition function} \cite{Dcalc}: this is a symmetric function $Z(u,v) = Z(v,u)$ of two non-intersecting simple loops (Jordan curves) $u$ and $v$ on $\hC$. It is a ratio of partition functions involving domains $U$ and $V$ bounded by $u$ and $v$ respectively; choosing them such that $u\subset V$ and $v\subset U$ (equivalently $\hC\setminus V\subset U$, or $\hC\setminus U\subset V$), and denoting $Z_A$ the CFT partition function on the domain $A$, it is defined (up to a $u$ and $v$-independent arbitrary normalization) as
\beq
	Z(u,v) := \frc{Z_U Z_V}{Z_{U\cap V}},\quad u=\p U,\;v=\p V,\quad
	u\subset V,\;v\subset U.
\eeq
We also define $Z(\emptyset,v) = Z(v,\emptyset) = 1$.

Suppose $\Or(x)$ is a local field in CFT, and denote its conformal transform as $g\cdot\Or(x)$; for instance, for a primary field of dimensions $(h,\t h)$, we have $g\cdot\Or(x) = (\p g(x))^h \,(\b \p \b g(\b x))^{\t h}\, \Or(g(x))$. Then the relation found in \cite{Dcalc}, encoding both the conformal Ward identities and the conformal boundary conditions on simply connected domains $A$, is
\beq\label{genwardCFT}
	\bra T(w) \Or(x)\cdots\ket_A
	= Z(\p A,v)^{-1} 
	\frc{\p}{\p \eta} \Big(Z(g_{w,\eta}(\p A),g_{w,\eta}(v))
	\;\bra g_{w,\eta}\cdot \Or(x)\cdots
	\ket_{g_{w,\eta}\cdot A} \Big)_{\eta=0}
\eeq
for $w\in A$, where $v\subset A$ is any simple loop that surrounds $w$. The fields hidden in the dots ``$\cdots$'' on the right-hand side are also affected by the $g_{w,\eta}$ transformation. The $\eta$-derivative is the usual complex derivative.

In \eqref{genwardCFT}, of course, since $w$ lies in the domain $A$, the map $g_{w,\eta}$ is not conformal on $A$ for any $\eta\neq0$. Yet, since $g_{w,\eta}$ is conformal at any point different from $w$ for $\eta$ small enough, then $g_{w,\eta}\cdot \Or(x)$ makes sense by using the explicit transformation properties of the CFT local operator $\Or(x)$. However, we must define what we mean by the domain $g_{w,\eta}\cdot A$. We simply define it as the domain bounded by $g_{w,\eta}(\p A)$ such that neighborhoods of $g_{w,\eta}(\p A)$ inside $g_{w,\eta}\cdot A$ are images of neighborhoods of $\p A$ inside $A$. This uniquely defines $g_{w,\eta}\cdot A$ for any domain $A$ containing the point $w$, and for any $\eta$ small enough.

As a simple check, take for instance $A=\hC$, in which case we have $g_{w,\eta}\cdot \hC = \hC$, and take $\Or(x)$ to be primary of dimensions $(h,\t h)$. Then
\[
	g_{w,\eta}\cdot \Or(x) = \lt(1+\frc{\eta}{(w-x)^2}\rt)^h\,
	\lt(1+\frc{\b \eta}{(\b w-\b x)^2}\rt)^{\t h} \Or\lt(x + \frc{\eta}{w-x}\rt)
\]
and calculating the $\eta$-derivative $\p \big(g_{w,\eta}\cdot \Or(x)\big) / \p \eta$ at $\eta=0$, we immediately find $\big(h/(w-x)^2 + 1/(w-x) \;\p_x\big) \Or(x)$, which is indeed the singular part of the $T(w)\Or(x)$ OPE. On $\hC$ this singular part fully determines the correlation function as a function of $w$, in agreement with \eqref{genwardCFT}.

It is \eqref{genwardCFT} that was shown in \cite{DTCLE} to hold for the variable $\tT(w)$ in CLE. That is, if $\tX$ is a CLE variable supported away from $w$, and both the support of $\tX$ and the point $w$ lie inside the domain $A$, then it was shown that
\beq\label{genwardCLE}
	\mathbb{E}\big[\tT(w) \,\tX\big]_A = Z(\p A,v)^{-1} 
	\frc{\p}{\p \eta} \Big(Z(g_{w,\eta}(\p A),g_{w,\eta}(v))\;
	\mathbb{E}\big[g_{w,\eta}\cdot \tX\big]_{g_{w,\eta}\cdot A} \Big)_{\eta=0}.
\eeq
In addition, it was also shown that the relative partition function $Z(u,v)$ has the following expression in CLE:
\beq
	Z(u,v) = \frc1{\mathbb{E}\big[\tE(u)\big]_V},\quad
	v=\p V,\quad
	u\subset  V.
\eeq
The relation between the CFT and CLE relative partition functions is that which allows to identify the number $c$ involved in \eqref{transT} with the CFT central charge \cite{Dcalc}.

This immediately implies the standard CFT results. For instance, let $\tX:=\tX(x)$ be a CLE variable that is supported on the point $x$ and that transforms like a primary field, $g\cdot \tX(x) = (\p g(x))^h (\b \p \b g(\b x))^{\t h}\, \tX(g(x))$. This transformation property has the usual conformal-invariance meaning: $\mathbb{E}\big[g\cdot \tX(x) \, g\cdot \tX'\big]_{g(A)} = \mathbb{E}\big[\tX(x) \, \tX'\big]_{A}$ for any $g$ conformal on $A$ (any domain which includes the point $x$ and the support of $\tX'$) and any $\tX'$ supported away from $x$. Then the OPE between $\tT(w)$ and $\tX(x)$ is
\beq
	\tT(w) \tX(x) = \frc{h}{(w-x)^2} \tX(x) + \frc1{w-x} \frc{\p}{\p x} \tX(x) 
	+ \ldots.
\eeq
One can also, of course, define the anti-holomorphic stress-energy tensor variable $\b \tT(\b w)$ as the spin-$(-2)$ rotating ellipse:
\beq
	\b \tT(w) = \lim_{\ell\to0} \frc1{2\pi \ep^2}\int_0^{2\pi} d\theta\,e^{2i\theta} \,\tE(\alpha(w,\theta,\ell,e)),\quad \ep = \frc{e\ell}2.
\eeq
Then, it turns out, Eq. \eqref{genwardCLE} and its equivalent for $\b \tT(\b w)$ also implies that the relation
\beq
	\lim_{{\rm Im}(w)\to0} (\tT(w) - \b\tT(w)) = 0
\eeq
holds on the upper half plane $\uH$, or on any domain with a boundary component that is the real line. This is the usual conformal boundary condition on the real line \cite{C84}. Along with the OPE's for $\b \tT(\b w)$ and the transformation property \eqref{transT}, and using Riemann's mapping theorem, this then uniquely fixes the $w$-dependence of the insertion of $\tT(w)$ in any expectation value (with other variables supported away from $w$) on any simply connected domain (containing $w$), in agreement with the usual CFT results.

However, it is worth noting that the result \eqref{genwardCLE} contains more than the standard CFT formulas, as it can be applied to {\em any} CLE variable $\tX$ supported in $A\setminus\{w\}$. This includes variables like the number of loops that intersect simultaneously two disjoint domains, etc. Such variables are generically non-local (supported on a continuum instead of a set of points). Hence, \eqref{genwardCLE} should be interpreted as containing a {\em generalization of the conformal Ward identities to non-local observables}. 

All these results are quite nontrivial statements about CLE variables, and are essentially enough to fully identify the variable $\tT(w)$ with the CFT stress-energy tensor. 

\begin{rema}
As was observed in \cite{Dcalc}, the derivative with respect to small variations of the domain boundary $\partial A$ in \eqref{genwardCFT} is an indication that the domain boundary may be interpreted as a continuum of zero-dimensional primary fields. In this interpretation, the conformal boundary conditions are then just the usual conformal Ward identities applied to such a continuum. On the other hand, the part that involves the derivative of the relative partition function is just the expectation value of the stress-energy tensor,
\beq\label{onept}
	\bra T(w)\ket_A = \frc{\p}{\p \eta} \log Z(g_{w,\eta}(\p A),g_{w,\eta}(v)).
\eeq
\end{rema}

\begin{rema}
Relation \eqref{genwardCLE} was also generalized to multiply connected domains in \cite{Ddesc}, where the only change is in the definition of the relative partition function: let $U_i,\,i=1,\ldots,n$ be Jordan domains with pairwise disjoint complements, let $U = U_1\cap U_2\cap \cdots\cap U_n$, $u_i = \p U_i$ and $u =\p U$, and let $v\subset U$ be a Jordain curve, then
\beq
	Z(u,v) := \frc{\mathbb{E}\big[\prod_i \tE(u_i)]_\hC}{
	\mathbb{E}\big[\prod_i \tE(u_i)]_V},\quad v=\p V,\quad u\subset V.
\eeq
\end{rema}

\section{Hypotrochoids and the Virasoro algebra}

\subsection{Hypotrochoids variables}

In the previous section, we interpreted the stress-energy tensor two-point function \eqref{2pt} as the simplest shape correlation, and as indicating that some ``intensity'' of transfer of fluctuations from small to large scales is proportional to $c$. The fluctuations were seen as being 2-crest fluctuations traveling with spin 2 on a small circle. Are there results for more general types of fluctuations?

\begin{figure}\bc
\includegraphics[width=3 cm]{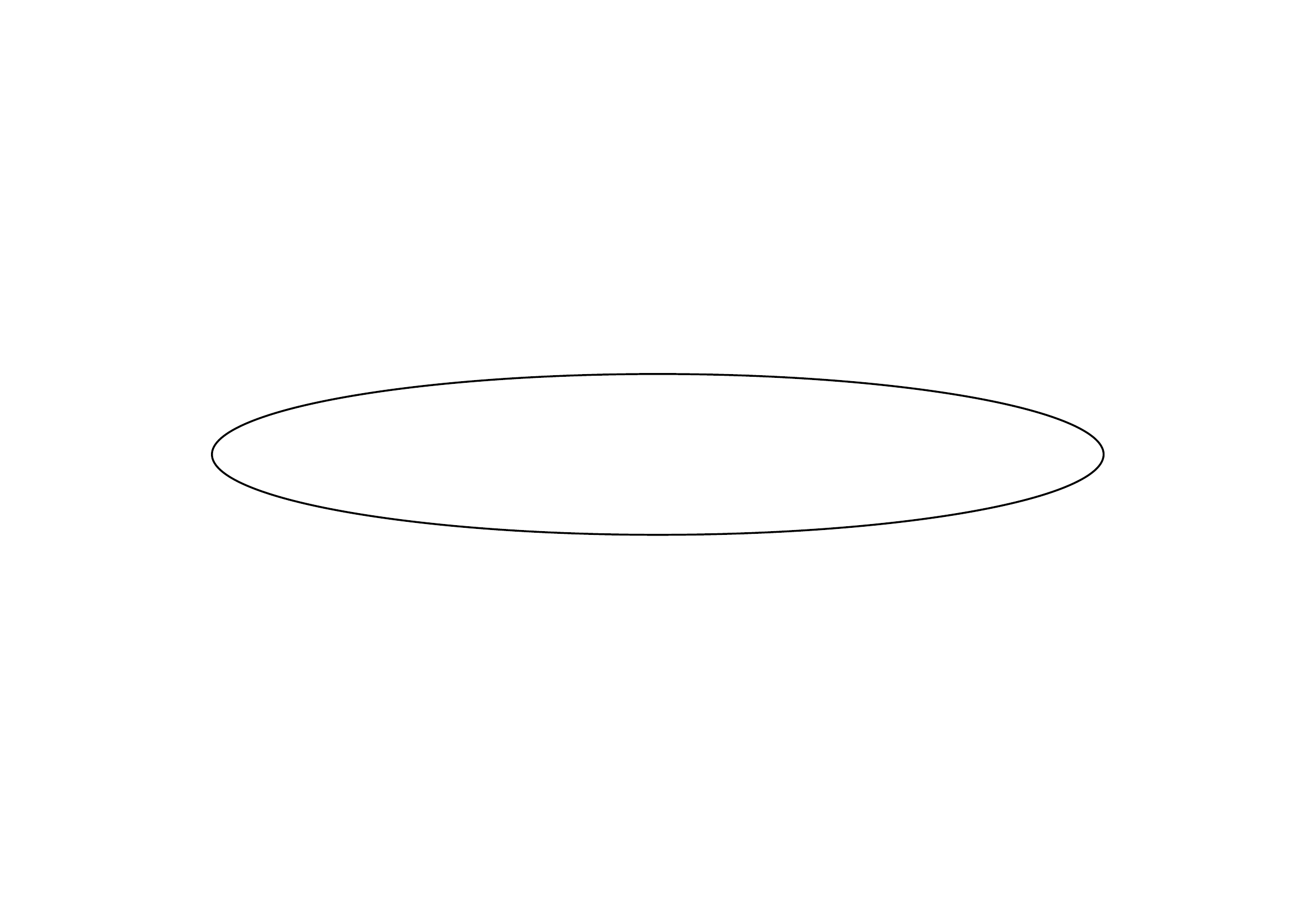} 
\includegraphics[width=3 cm]{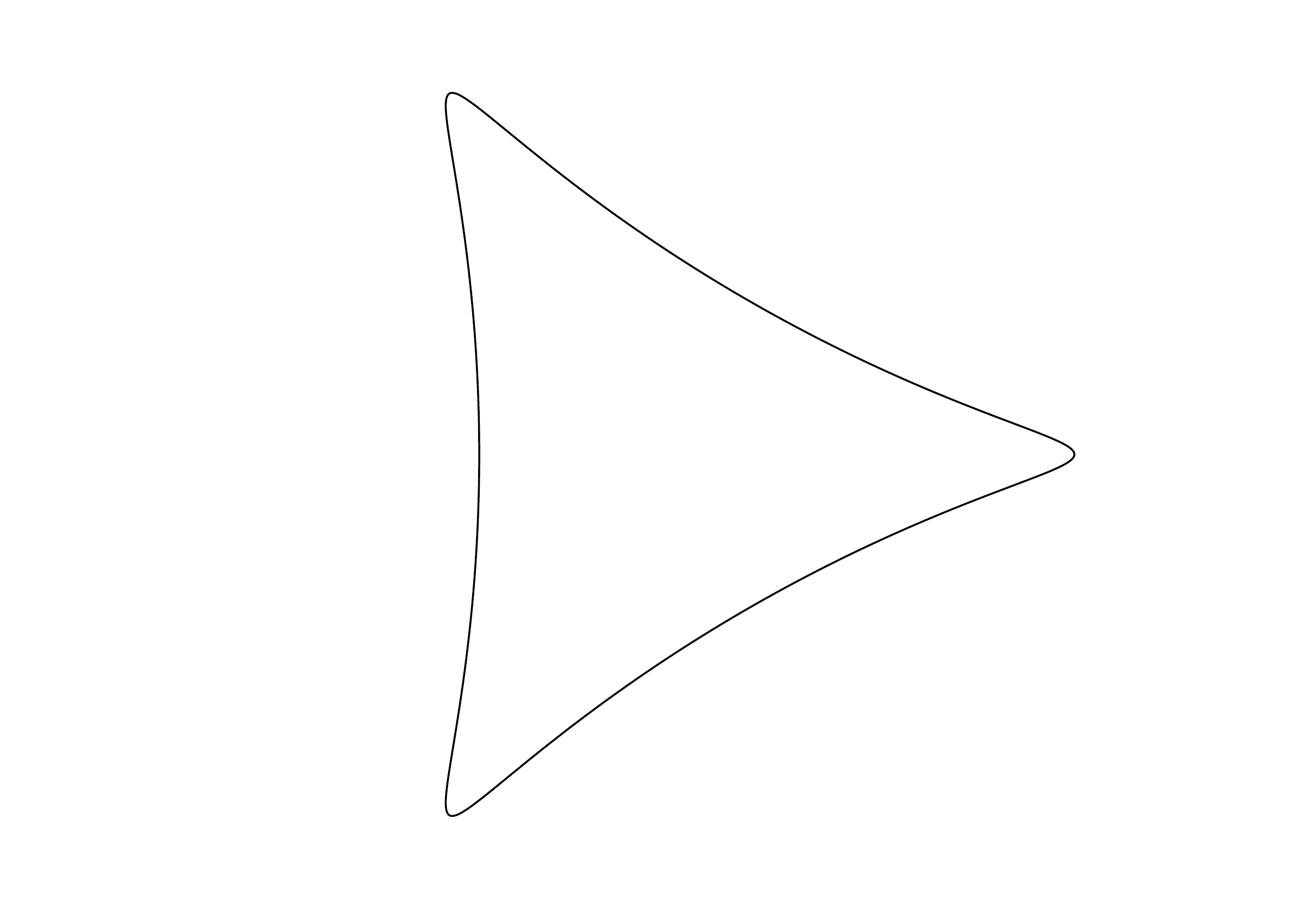} 
\includegraphics[width=3 cm]{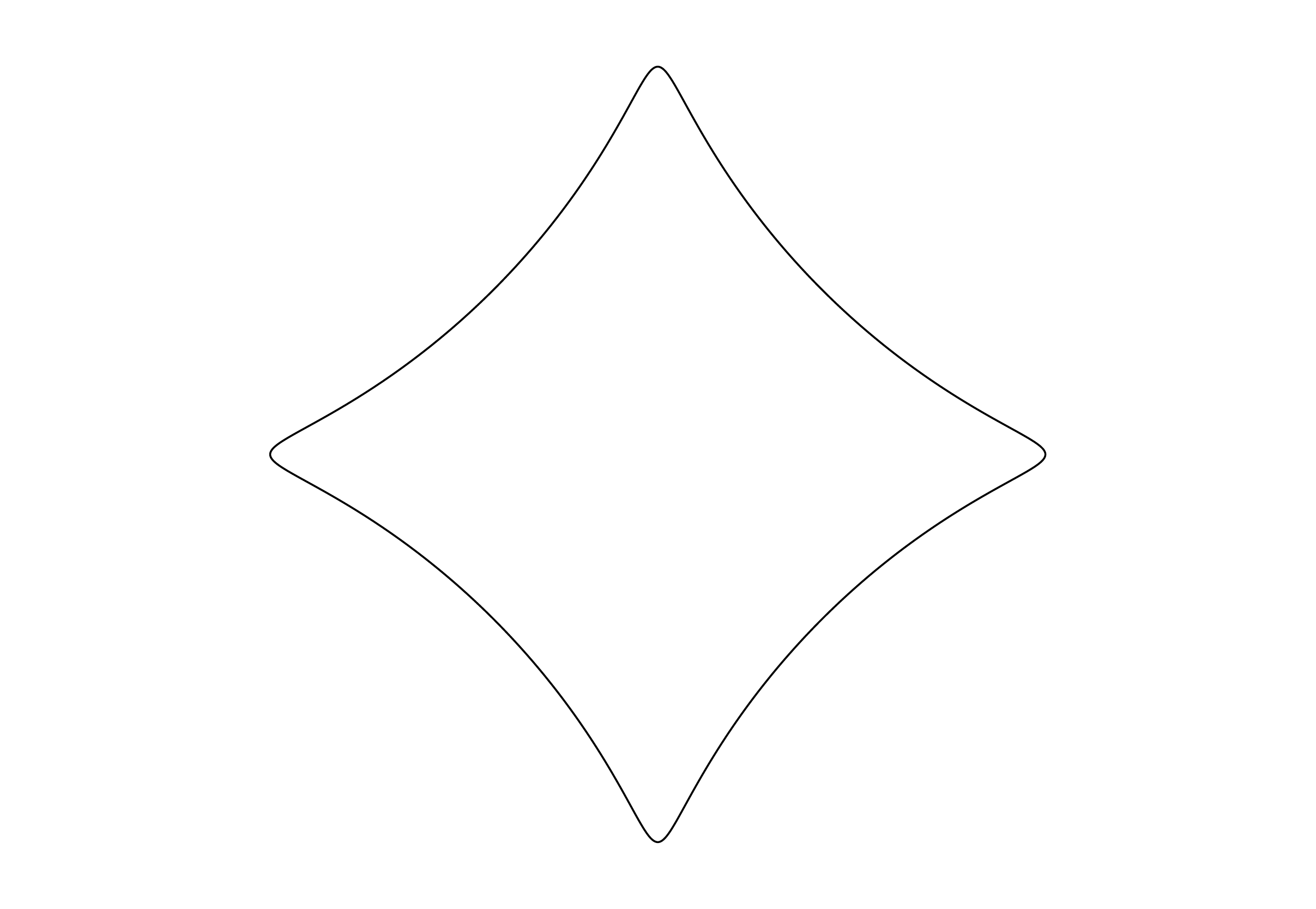} 
\includegraphics[width=3 cm]{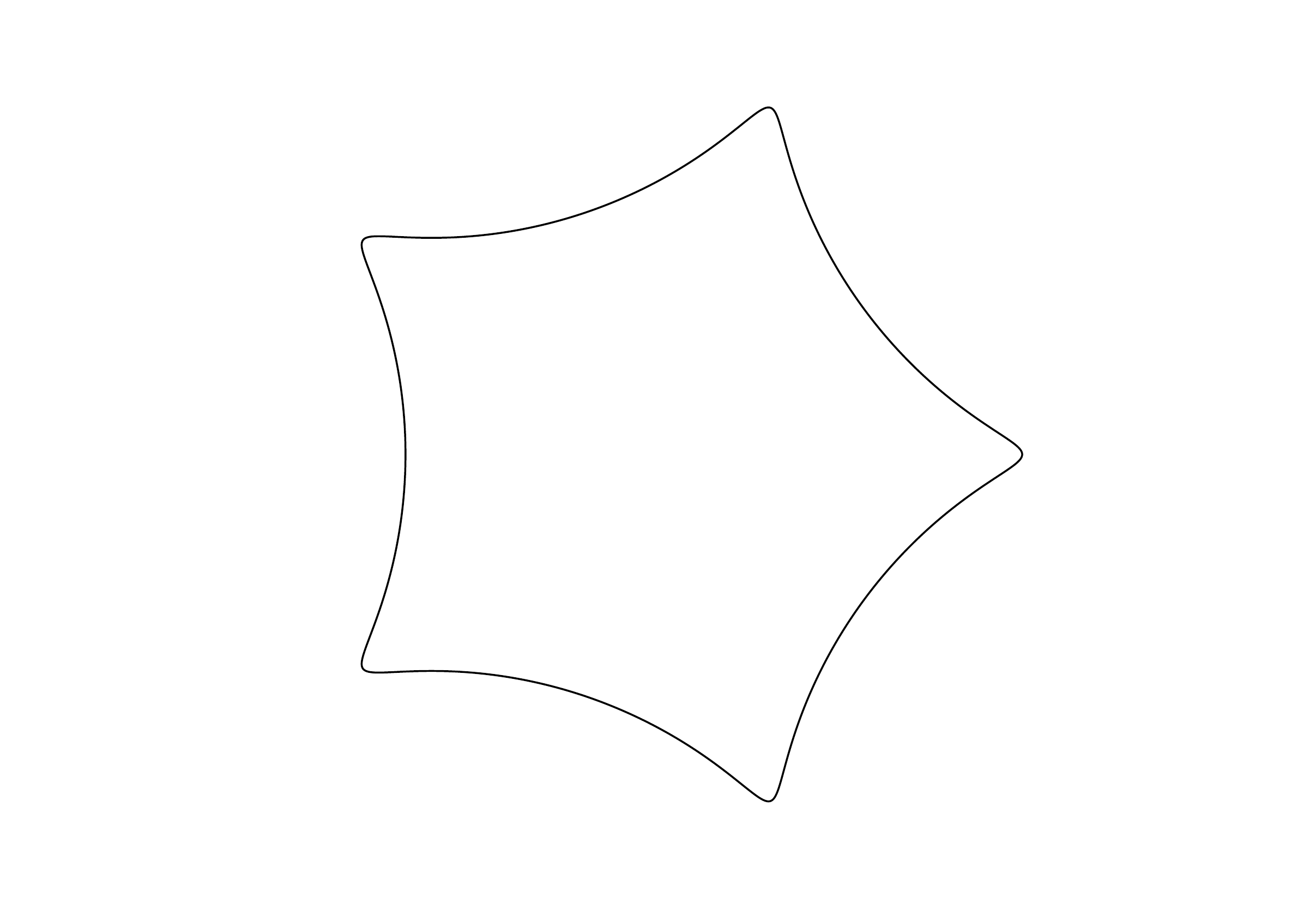} \ec
\caption{Hypotrochoids, with $k=2$, $k=3$, $k=4$ and $k=5$.}\label{hypo}
\end{figure}

It turns out that we do have results for higher spins, and also for a convenient ``basis'' of shapes with higher numbers of crests: the hypotrochoids. The hypotrochoids simple curves are natural generalizations of the ellipse, formed by tracing a fixed point in a disk as it rolls inside a bigger disk, with the condition, on the radii, that after one revolution the point is back to its initial position. Explicitly, the $k$-cycle hypotrochoid, which has exactly $k$ crests, is
\beq
	\alpha_k(w,\theta,\epsilon,b):=\left\{ w + \epsilon e^{\ii\theta} (b e^{\ii\beta} + b^{1-k} e^{(1-k)\ii\beta}):\beta\in[0,2\pi)\right\},
\eeq
for $b >  (k-1)^{1/k}$ (see Figure \ref{hypo}). The ellipse of the previous section, centered are $w$ with major semi-axis $\ell$ at angle $\theta$ from the positive real axis and with eccentricity $e$, is the case $k=2$ with $\ep=e \ell/2$ and $b=1/e + \sqrt{1/e^2-1}$. The case $b=(k-1)^{1/k}$ is the case of the hypocycloids, which have spikes instead of crests (for $k=2$, the ellipse degenerates to a segment).

We now consider the shape-measuring variable $\tE(\alpha_k(w,\theta,\ep,b))$ for the $k$-cycle hypotrochoid, and look at its Fourier transform in $\theta$, a function of the spin. By the symmetries of the hypotrochoids, it is clear that $\tE(\alpha_k(w,\theta,\ep,b))$ is periodic under $\theta\mapsto \theta+2\pi/k$. Hence we may restrict ourselves to spins $km$ with $m\in\Z$.

For the same reasons as in the previous section, the spin-$km$ Fourier component is expected to be zero as $\ep\to0$: loops surrounding the point $w$ shields it from the outside so that it is effectively on $\hC$, and by rotation symmetry the Fourier components vanish on $\hC$. But again, the way it tends to zero is simply proportionally to $\ep^{km}$, and we can define the variables \cite{Ddesc} (weakly locally):
\beq\label{Tkmdef}
	\tT_{k,m}(w) = \lim_{\ep\to0} \frc{m!}{2\pi \ep^{km}}\int_0^{2\pi} d\theta\,e^{-km \ii \theta} \,\tE(\alpha_k(w,\theta,\ep,b)).
\eeq
The case \eqref{T21} is $k=2$ and $m=1$. A similar definition can be made for $\b \tT_{k,m}(\b w)$, simply by taking the complex conjugate of the right-hand side. In fact, we can gather all the $\tT_{k,m}(w)$ into a single asymptotic expansion of the shape-measuring variable itself, in terms of the complex variables
\[
	u = \ep e^{\ii\theta}\quad \mbox{and}\quad \b u = \ep^{-\ii\theta}.
\]
We obtain \cite{Ddesc}:
\beq\label{tEexp}
	\tE(\alpha_k(w,\theta,\ep,b)) = 1 + o(\ep^0) + \sum_{m=1}^\infty
	\frc{u^{km}}{m!}\lt( \tT_{k,m}(w) + o(\ep^0)\rt) + \sum_{m=1}^\infty
	\frc{{\b u}^{km}}{m!}\lt( \b \tT_{k,m}(w) + o(\ep^0)\rt).
\eeq

The three nontrivial statements in the expansion \eqref{tEexp} (and equivalently \eqref{Tkmdef}) are that: (i) the asymptotic expansion is dominated by series in non-negative integer powers of $u^k$ and $\b u^k$, (ii) the coefficients in these series are holomorphic and antihomorphic in $w$, respectively, and (iii) the coefficients in these series are independent of the parameter $b$ (all statements being true weakly locally). The latter fact says that many different shapes give rise, in the appropriate limit, to the same variable -- this is one aspect of a large universality, about which we will not comment but which was discussed in \cite{DTCLE}. The first fact gives us another way of defining the variables $\tT_{k,m}(w)$ and $\b\tT_{k,m}(\b w)$, in terms of complex derivatives with respect to $u^k$ and $\b u^k$:
\beq
	\tT_{k,m}(w) = \lt(\frc{\p}{\p u^k}\rt)^m \tE(\alpha_k(w,\theta,\ep,b))\Big|_{\ep=0},\quad
	\b \tT_{k,m}(\b w) = \lt(\frc{\p}{\p \b u^k}\rt)^m \tE(\alpha_k(w,\theta,\ep,b))\Big|_{\ep=0}.
\eeq
We have here a nice relation between complex derivatives in $u^k$ and $\b u^k$, and holomorphicity in $w$ and $\b w$. Formally (i.e. without worrying about convergence) we may extract the holomorphic and antiholomorphic parts as new variables:
\beq\label{holoexp}
	\tE_k(w,u) := \sum_{m=1}^\infty \frc{u^{km}}{m!} \tT_{k,m}(w),\quad
	\b\tE_k(\b w, \b u) := \sum_{m=1}^\infty \frc{\b u^{km}}{m!} \b \tT_{k,m}(w)
\eeq
so that $\tE(\alpha_k(w,\theta,\ep,b)) = 1 + \tE_k(w,u) + \b \tE_k(\b w, \b u) + \cdots$.

Can we calculate expectations containing the variables $\tT_{k,m}(w)$? It turns out that indeed all expectations can be calculated in principle, and that these have exactly the structure of correlation functions of {\em Virasoro descendants} of the holomorphic stress-energy tensor of CFT. That is, there are fields in CFT, obtained via it Virasoro algebraic structure and which we denote $T_{k,m}(w)$, such that we may make the identification $\tT_{k,m}(w) \equiv T_{k,m}(w)$:
\beq\label{expTkm}
	\mathbb{E}\big[\prod_i \tT_{k_i,m_i}(w_i) \,\tX\big]_A
	= \bra \prod_i T_{k_i,m_i}(w_i)\,\Or_{\tX}\ket_A.
\eeq
On the left-hand side, we've put an arbitrary random variable $\tX$, which should be supported in $A$ but away from $\{w_1,w_2,\ldots\}$. On the right-hand side, we have a CFT correlation function, with the CFT operator $\Or_\tX$ corresponding to the random variable $\tX$ (it is just the identity if $\tX$ is the identity). If the random variable $\tX$ is supported on a point $x$ or a set of points $x_1,x_2,\ldots$, we would expect $\Or_\tX$ to be a local field or a product thereof, at the points $x_1,x_2,\ldots$. Otherwise, it is a more complicated operator. If we know what $\Or_\tX$ is, then we've reduced the calculation of a nontrivial expectation value in CLE to the calculation of a correlation function in CFT.

But in fact, we don't need to know much about what $\Or_\tX$ actually is in order to evaluate the CFT correlation functions, at least on $\hC$ or on any simply connected domain. Indeed, thanks to the conformal Ward identities and the conformal boundary conditions for $T_{k,m}(w)$, all we need to know is it's transformation properties under conformal maps, and its averages $\bra \Or_\tX\ket_A = \mathbb{E}\big[\tX\big]_A$.

In order to express all this more precisely, let us discuss first the conformal descendants $T_{k,m}$. Without going into the full details of what these fields are, they are descendants on the form
\beq\label{Tkm}
	T_{k,m} = \sum_{\lambda
	= (\lambda_1,\ldots,\lambda_j)
	\atop \lambda\in\Phi(m)} C_\lambda (k-1)^{m-j}
	L_{-k\lambda_j}\cdots L_{-k\lambda_2} L_{-k\lambda_1} {\bf 1}
\eeq
where $\Phi(m)$ is the set of all ordered partitions of $m$ (if $\lambda\in\Phi(m)$ then $\sum_{i=1}^j \lambda_i = m$ and $\lambda_i\geq 1$, with $j\geq 1$). An explicit recursion relation for $C_\lambda$ is presented in \cite{Ddesc}. Special cases are
\beqa
	T_{k,1} &=& L_{-k}{\bf 1} \n
	T_{k,2} &=& (L_{-k}^2 + (k-1)L_{-2k}) {\bf 1} \n
	T_{k,3} &=& (L_{-k}^3 + 3(k-1) L_{-2k}L_{-k} + 2(k-1)(2k-1) L_{-3k}){\bf 1}.
\eeqa

Then, generalizing what we saw in the previous section, the extended conformal Ward identities (combination of the conformal Ward identities and the boundary conditions) for the descendants $T_{k,m}(w)$ can then be recast into the calculation of \eqref{expTkm} as multiple derivatives, conjugated by the relative partition function, with respect to small conformal maps (that are singular at the points $w_i$) of the average $\mathbb{E}\big[\tX\big]_A$. That is,
\beq\label{ETX}
	\mathbb{E}\big[\tT_{k,m}(w)\, \tX\big]_A =
	\sum_{\lambda= (\lambda_1,\ldots,\lambda_j)
	\atop \lambda\in\Phi(m)} C_\lambda (k-1)^{m-j}\;
	Z(\p A,v)^{-1} \Delta[h_{-k\lambda_j,w}]
	\cdots \Delta[h_{-k\lambda_1,w}]\,\Big(
	Z(\p A,v)\,\mathbb{E}\big[\tX\big]_A\Big)
\eeq
where again $v\subset A$ is any Jordan curve surrounding $w\in A$. Here $h_{\ell,w}(z) = -(z-w)^{\ell+1}$ are holomorphic functions in $\hC\setminus \{w\}$ which determine the ``direction'' in which the small conformal transformation is taken. The notation $\Delta[h] f$, for a function $f$, means the complex derivative with respect to $\eta$ of the function obtained by acting with the small conformal map $g(z) = z + \eta h(z)$ on the argument of $f$:
\beq\label{Delta}
	\Delta[h] f(-) = \frc{\p}{\p\eta} f\big((\id + \eta h)\cdot -\big).
\eeq
In our case, the function $f(\tX,\p A,v)$ is $Z(\p A,v)\,\mathbb{E}\big[\tX\big]_A$ where the argument is the triple $(\tX,\p A,v)$, and the action is $g\cdot (\tX,\p A,v) = (g\cdot \tX, g(\p A), g(v))$.

\begin{figure}\bc
\includegraphics[width=12 cm]{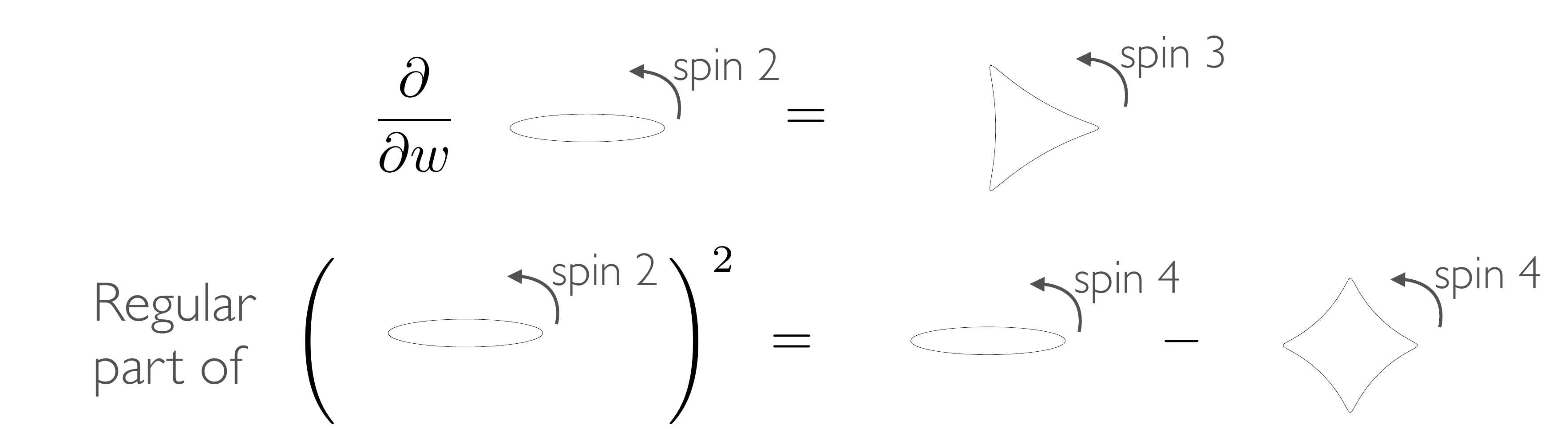} \ec
\caption{Graphical representations of the relations \eqref{rel1} and \eqref{rel2} obtained from the Virasoro algebra structure. The shapes are centered at the point $w$.}\label{OPE}
\end{figure}

The identification with descendants of the stress-energy tensor allows more than to calculate CLE expectations in terms of CFT correlation functions. Indeed, the Virasoro vertex operator algebraic structure underlying Wilson's operator product expansion of the CFT fields involved give rise to nontrivial relations amongst our CLE random variables $\tT_{k,m}(w)$. For instance, we find
\beq\label{rel1}
	\frc{\p}{\p w} \tT_{k,1}(w) = \tT_{k+1,1}(w).
\eeq
That is, differentiating with respect to $w$ the $k$-crest, spin-$k$ wave gives the $(k+1)$-crest, spin-$(k+1)$ wave. There are also relations in order to evaluate finite parts of products, for instance
\beq\label{rel2}
	\lt[\tT_{2,1}(w) \tT_{2,1}(w')-\mbox{singular part in $(w-w')$}\rt]_{w=w'}
	= \tT_{2,2}(w) - \tT_{4,1}(w)
\eeq
and many other similar relations, not easy to guess from CLE but consequences of the Virasoro algebra (see Figure \ref{OPE}).

\begin{rema}
We note that \eqref{tEexp} looks very much like part of an ``operator product expansion'' for the variable $\tE(\alpha_k(w,\theta,\ep,b))$ itself. The number 1 and the holomorphic and anti-holomorphic parts $\tE(w,u)$ and $\b \tE(\b w, \b u)$ can be identified with the identity sector of the OPE. In fact, \eqref{tEexp} is the normalized operator product expansion, in the variable $u$, of the product $\prod_{j=0}^{k-1}{\cal T}_2(w+u e^{2\pi \ii j /k})$, where ${\cal T}_2(z)$ is the {\em branch-point twist field} for a square-root branch point at $z$. Branch point twist fields for $n^{\rm th}$ root branch points are twist fields (local in the QFT sense) associated to the cyclic permutation symmetry of a $n$-copy version of the model, and geometrically correspond to inserting branch points, or infinite-negative-curvature points with angle $2\pi n$. To our knowledge, they were first described in CFT in \cite{K87} (referred to as ``analytic fields'') and in general QFT in \cite{CCD08}, and have found very fruitful recent applications in the context of entanglement entropy starting from the work \cite{CC04}.
\end{rema}

\begin{rema}
As we alluded to, we may again interpret our results \eqref{expTkm} as indicative of correlations amongst certain fluctuations traveling around small circles that are far apart, and between such fluctuations and large-scale objects. Here, the fluctuations are $k$-crest fluctuations traveling with a spin (rotational velocity) $km$. Hence we find that correlations amongst such fluctuations generate part of the Virasoro vertex operator algebra structure. However, they do not generate the whole Virasoro vertex operator algebra, and the fluctuation interpretation of all Virasoro descendants has not yet been elucidated.
\end{rema}

\begin{rema}
Although the stress-energy tensor has a clear physical interpretation in CFT when viewed as a {\em quantum} field theory, it is not as clear when viewed as a {\em statistical} field theory. The physical interpretation for its Virasoro descendants are even less clear; they were introduced in \cite{BPZ} as algebraic constructions only. Here we have for the first time a {\em statistical interpretation} of the stress-energy tensor along with a large family of its descendants.
\end{rema}

\subsection{Conformal geometry}

Comparing \eqref{ETX} and \eqref{Tkm} we see that the extended conformal Ward identities have a natural interpretation in terms of representation of the Virasoro algebra as certain differential operators. Indeed, \eqref{ETX} is obtained from \eqref{Tkm} essentially by making the substitution
\beq
	L_n \mapsto Z(\p A,v)^{-1} \Delta[h_{n,w}] Z(\p A,v).
\eeq
This is a particular case of a general theory of representation of the Virasoro algebra, and of the Virasoro vertex operator algebra, on spaces of functions of conformal maps.

The Virasoro algebra is the central extension of the Witt algebra. The Witt algebra is generated by $-z^{n+1} \p_z$, and these differential operators can be seen as vector fields on some annular domain that excludes the points $0$ and $\infty$. This gives a representation of the Witt algebra on the space of holomorphic functions on this domain: vector fields naturally act on these functions. But we may take a different viewpoint. The space of vector fields on a given annular domain $E$ can also be interpreted as the tangent space at the identity, of a variety $\Omega(E)$ of conformal maps on the domain. Indeed, small conformal transformations $g = \id + \eta h$ are characterized by holomorphic functions $h$. Conformal transformation properties agree with the identification of $h$ with vector fields $h \p$, and conformal map composition gives rise to the vector field Lie algebra. In this viewpoint, one considers functions not on $E$, but rather on $\Omega(E)$; functions which take conformal maps as arguments, and give complex numbers, say. The tangent space at the identity is the space of derivatives of such functions at the identity, identified with the space of holomorphic vector fields. One can check that derivatives $\Delta [h]$, defined in \eqref{Delta}, indeed satisfy the commutation relations of holomorphic vector fields,
\beq
	[\Delta[h],\Delta[h']] = \Delta[h\p h' - h'\p h].
\eeq
This gives a representation of the Witt algebra on a much bigger space, that of functions of conformal maps. Since conformal maps form a groupoid, one can construct a structure similar to that of Lie groups on $\Omega(E)$, so that the tangent space at the identity is sufficient to describe the tangent space at other points. Hence, although the space $\Omega(E)$ is much bigger, the algebra of vector fields on $\Omega(E)$ is reduced to the ``Lie algebra'' associated to the ``Lie group'' $\Omega(E)$, which is isomorphic to the algebra of vector fields on $E$.

But the room afforded by considering the bigger space $\Omega(E)$ instead of $E$ allows us to represent the Virasoro algebra on the same space. The idea is to provide a nontrivial {\em connection} on that space, by defining the ``covariant derivative''
\beq\label{Dh}
	D[h] := \Delta[h] + \Gamma[h].
\eeq
The connection $\Gamma[h]$ is a function on the space, which depends on the direction in which we differentiate, $h$. A connection that gives rise to the Virasoro algebra can be defined on the basis $h_{n}(z):= -z^{n+1}$ ($n\in\Z$) as follows:
\beq\label{connect}
	\Gamma[h_{n}] := \lt\{\ba{ll}
	\Delta[h_{n}] \log Z & (n\leq -2) \\ 0 & (n\geq -1) \ea\rt.
\eeq
where $Z$ is a function that is only required to satisfy the following second order differential equations:
\beq\label{diffdelta}
	\lt. \ba{ll}
	\Delta[h_n] \Delta[h_m] \log Z & (n+m\geq -1) \\
	\Delta[h_m] \Delta[h_n] \log Z & (n+m\leq -2) \ea \rt\}
	= \frc{c}{12} (n^3-n)\delta_{n+m,0}.
\eeq
Reinterpreting \eqref{Dh} and \eqref{connect}, we see that  $D[h_n]$ is just the original derivative $\Delta[h_n]$ for $n\geq -1$, but it is its conjugate $D[h_n] = Z^{-1} \Delta[h_n] Z$ for $n\leq -2$. This representation leads to the Virasoro algebra with central charge $c$:
\beq
	[D[h_n],D[h_m]] = (n-m) D[h_{n+m}] + \frc{c}{12} (n^3-n) \delta_{n+m,0}.
\eeq
In this picture, the central charge is seen as providing a nonzero curvature ${\cal R}_{n,m} := [D[h_n],D[h_m]] - D[h_n\p h_m - h_m\p h_n]$:
\beq
	{\cal R}_{n,m} = \frc{c}{12} (n^3-n) \delta_{n+m,0}.
\eeq

In the context of CLE or CFT, the function $Z$, as seen in the previous subsection, is just the relative partition function $Z(\p A,v)$, and it turns out that it indeed satisfies the differential equations \eqref{diffdelta} \cite{Dhigher}. This provides a geometric interpretation for the differential operators involved in the extended Ward identities of CFT. This construction was extended in \cite{Dhigher} to a construction of the full Virasoro vertex operator algebra \cite{LL04}, providing a reinterpretation of Huang's geometric vertex operator algebra \cite{Hu97} that is adapted to the mathematics of statistical field theory. In Huang's construction, correlation functions are the objects on which the representation of the Virasoro algebra acts, and we have extended this to an action on CLE expectation functions. Connections between CFT and SLE via related geometric constructions were also made in \cite{Fried1,Fried2}. However, the above representation and geometric interpretation is quite general, and indeed one expects that vertex operator algebra structures can be extracted in general situations not immediately related to CFT correlation functions or CLE expectations.

\section{From small to large scales}

We now provide a more speculative discussion about the fluctuation interpretation of some of the above results.

Let us go back to the small spin-2 rotating ellipse, and consider formula \eqref{genwardCLE}. For definiteness let us take the special case $\tX = \tI(N)$ and $A=\hC$: the indicator variable $\tI(N)$ for there to be at least one loop in the annular domain $N$, on the Riemann sphere $\hC$. Using our basic definition \eqref{T21}, we can re-write this as
\beq
	\mathbb{E}\big[
	\int_0^{2\pi} \frc{d\theta}{2\pi}\,e^{-2\ii\theta}
	\,\tE(\alpha(w,\theta,\ell,e))
	 \,\tI(N)\big]_\hC = \ep^2\,
	\frc{\p}{\p\eta} \mathbb{E}\big[\tI(g_{w,\eta}(N))]_{\hC}\big|_{\eta=0}
	+ o(\ep^2)
\eeq
where $g_{w,\eta}$ is given by \eqref{gweta}. We observe that the right-hand side is proportional to $\ep^2$. With our usual interpretation of the Fourier transform of the variable $\tE(\alpha(w,\theta,\ell,e))$, this means that the correlation between the 2-crest, spin-2 small fluctuations, traveling on a circle of radius of order $\ep$, correlates with the macroscopic loops winding in $N$ with an intensity of order $\ep^2$. This tells us how much of the small fluctuations around a point travel to the larger loops: the 2-crest, spin-2 fluctuations, on loops of diameters of order $\ep$ and of amplitude of order $\ep$, spread out towards larger loops, so that there effects on macroscopic loops are of order $\ep^2$. We may understand this intuitively by the fact that as such fluctuations spread to loops bigger by a factor $\Lambda$, their amplitudes are reduced by a factor $\Lambda$, as if there was only a pure ``stretching'' effect.

\begin{floatingfigure}{4cm}\bc
\includegraphics[width=4 cm]{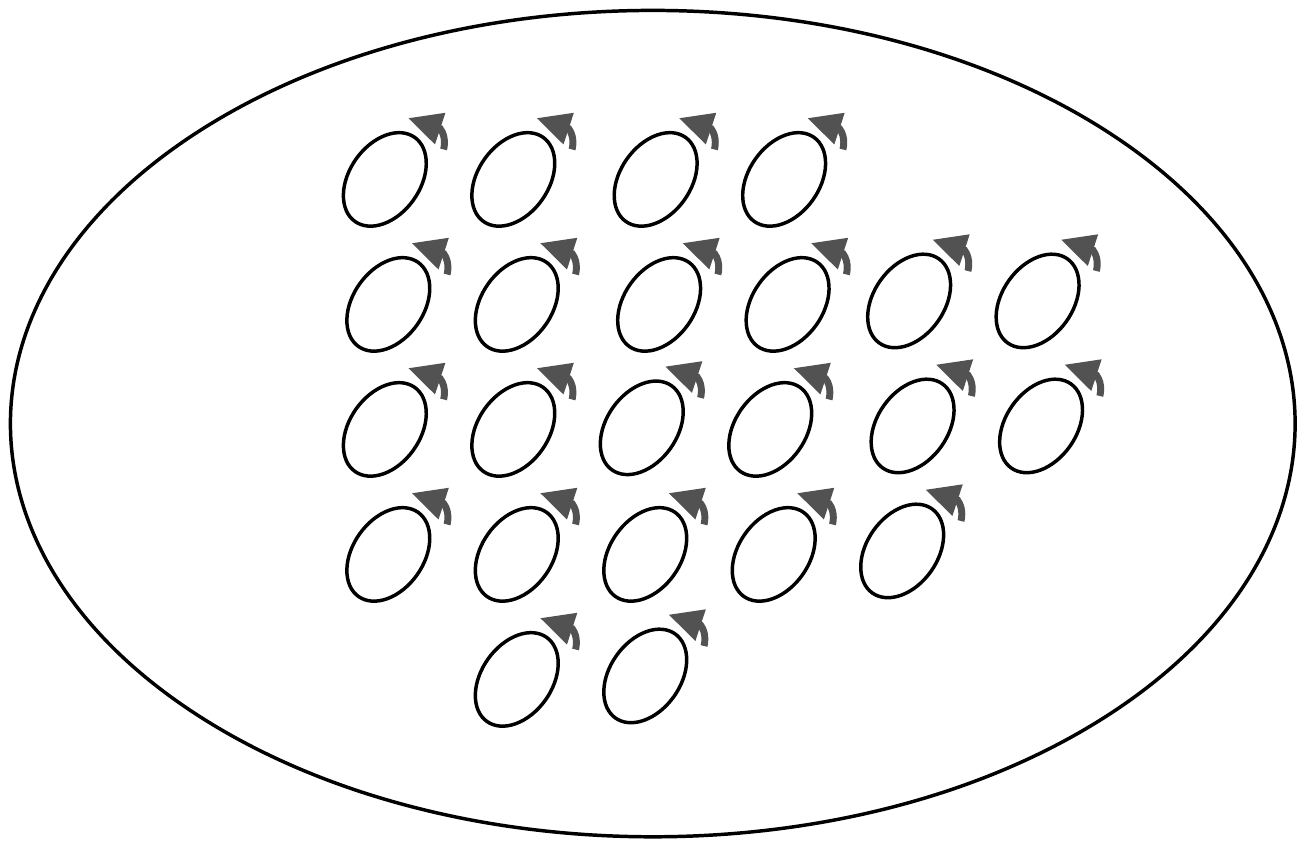} \ec
\caption{The cumulative effect of 2-crest, spin-2 small loop fluctuations propagating to large loops gives macroscopic fluctuations.}\label{smalllarge}
\end{floatingfigure}

The exact way that these small fluctuations correlate with large loops depends on what the variable $\frc{\p}{\p\eta} \tI(g_{w,\eta}(N)\big|_{\eta=0}$ is. We know from Subsection \ref{ssectshapecorr} that transfer of fluctuations should be proportional to the central charge $c$. Hence, we could guess that $\frc{\p}{\p\eta} \tI(g_{w,\eta}(N)\big|_{\eta=0}$ is related, in some way, to $\tI(N)$ with a factor proportional to the central charge. Unfortunately, there are no results yet for what expectation values of $\tI(N)$ are in CLE, so that we cannot say much more about the exact form of the fluctuation correlations between small and large scales.

Nevertheless, the $O(\ep^2)$ correlation is already an interesting conclusion. Let us imagine that we divide the region surrounded by the annular domain $N$ (in the finite complex plane) into $1/\ep^2$ small ``boxes'' of diameter of order $\ep$, and consider the small loops in these boxes (see Figure \ref{smalllarge}). For each of these boxes, we may ask about how fluctuations of loops there affect the macroscopic loops winding in $N$. From the above conclusion, the 2-crest, spin-2 fluctuations in each box should affect the macroscopic loop to order $\ep^2$. Since there are $1/\ep^2$ boxes, this means that the sum of each of such small local fluctuations produce $O(1)$, or macroscopic, fluctuations of large loops. In other words, it is the transfer of the 2-crest, spin-2 microscopic fluctuations distributed everywhere in space, from small to large scales, that determine the macroscopic fluctuations of large loops.

We  may of course generalize this to our $k$-crest, spin-$km$ fluctuations via the variable \eqref{Tkmdef}. With similar arguments, we find that such fluctuations affect macroscopic loops at order $\ep^{km}$. Hence, as these fluctuations travel towards loops larger by a factor $\Lambda$, their amplitude is decreased by a factor $\Lambda^{km-1}$: besides the ``stretching'' effect, there seems to be a ``loss'' by a factor $\Lambda^{km-2}$, with a ``loss exponent'' of $km-2$. As for the cumulative effect of the $k$-crest, spin-$km$ fluctuations on macroscopic loops, again by similar arguments, it is of order $\ep^{km-2}$ (involving the loss exponent). Hence, such fluctuations do not determine macroscopic fluctuations of large loops: they do not affect their macroscopic shapes. One may guess however that they are responsible for the {\em fractal structure} of large loops: the microscopic fluctuations of the large loops themselves, making them into very rough curves. For various values of $km$, one gets fluctuations at various, well-separated scales on the large loops.

A full investigation concerning how microscopic fluctuations give rise macroscopic fluctuations and fractal (or multi-fractal) structures would be very interesting.

\section{Conclusion}

In this paper I have reviewed some of the results concerning the relation between CLE and CFT. I have explained how these results tell us about the shape of loops and, in a sense, their fluctuations. This gives rise to an interpretation in terms of transfer of loop fluctuations between scales. The central charge is naturally interpreted as a measure of the flow of fluctuations, and somewhat precise statements can be made about how certain small-loop fluctuations propagate to large loops. I have also briefly reviewed aspects of the relation between the algebraic or probabilistic descriptions of correlation functions, and the form of the conformal Ward identities involving derivatives of functions on spaces of conformal maps. There, the central charge is interpreted as giving rise to a nonzero curvature induced by a connection that is simply related to partition functions in CFT, or to certain expectations in CLE. These conformal geometric structures provide adaptations, in the present context, of structures that appear in the theory of vertex operator algebras \cite{Hu97,Hu99,Hu03}, developed in order to make the relation between vertex operator algebras and Segal's construction of CFT.

I believe the random loop viewpoint clarifies many aspects of CFT. For instance, as argued here, it closely connects with the theory of nucleation and with various fundamental tenets of criticality. It also sheds light on the meaning of local conformal invariance and of the associated infinite conformal symmetry. Of course, many aspects of the works \cite{Dcalc,DTCLE,Ddesc,Dhigher} were not reviewed here. For instance, the conformal geometry description, which transcends CFT and CLE, suggests the potential for CFT-like structures in situations beyond the traditional ones; an attempt at a general formalism adapted to this was given in \cite{Ddesc} (conformal restriction systems).

It would be very interesting to further develop the ideas touched upon in this review.

{\bf Acknowledgments.}
I thank the Simons Center for Geometry and Physics for the hospitality and support while part of this work was done, and J. Cardy and A. Vas'eliev for discussions while there. I thank A. Recknagel and G. Watts for comments on the manuscript and discussions. I am also grateful to Seoul National University and the XXV IUPAP International Conference on Statistical Physics, where this work was presented. Finally, I acknowledge support of the EPSRC First Grant Scheme, project ``From conformal loop ensembles to conformal field theory'' EP/H051619/1, under which most of the results reviewed were obtained.

\end{document}